\documentclass[twocolumn]{aastex631}

\received{September 3, 2023}
\accepted{November 1, 2023}

\submitjournal{ApJ}

\usepackage[normalem]{ulem}
\usepackage{threeparttable}
\usepackage{amsmath}

\usepackage{xcolor}
\DeclareRobustCommand{\erase}{\bgroup\markoverwith{\textcolor{red}{\rule[.5ex]{2pt}{0.4pt}}}\ULon}

\usepackage{float}

\shorttitle{Double-peaked radio light curves in core-collapse supernovae}
\shortauthors{Matsuoka et al.}

\graphicspath{{./}{figure2/}}

\begin{document}

\title{A New Insight into Electron Acceleration Properties from Theoretical Modeling of Double-Peaked Radio Light Curves in Core-Collapse Supernovae}

\author[0000-0002-6916-3559]{Tomoki Matsuoka}
\correspondingauthor{Tomoki Matsuoka}
\email{tmatsuoka@asiaa.sinica.edu.tw}
\affiliation{Institute of Astronomy and Astrophysics, Academia Sinica, No.1, Sec.4, Roosevelt Road, Taipei 10617, Taiwan, R.O.C}
\affiliation{Department of Earth Science and Astronomy, Graduate School of Arts and Sciences, The University of Tokyo, Tokyo 153-8902, Japan}

\author[0000-0003-2579-7266]{Shigeo S. Kimura}
\affiliation{Frontier Research Institute for Interdisciplinary Sciences, Tohoku University, Aramaki-Aoba, Aoba-ku, Sendai 980-8578, Japan}
\affiliation{Astronomical Institute, Graduate School of Science, Tohoku University, Aramaki-Aoba, Aoba-ku, Sendai 980-8578, Japan}

\author[0000-0003-2611-7269]{Keiichi Maeda}
\affiliation{Department of Astronomy, Kyoto University, Kitashirakawa-oiwakecho, Sakyo-ku, Kyoto 606-8502, Japan}

\author[0000-0001-8253-6850]{Masaomi Tanaka}
\affiliation{Astronomical Institute, Graduate School of Science, Tohoku University, Aramaki-Aoba, Aoba-ku, Sendai 980-8578, Japan}

\begin{abstract}
It is recognized that some core-collapse supernovae (SNe) show a double-peaked radio light curve within a few years since the explosion.
A shell of circumstellar medium (CSM) detached from the SN progenitor has been considered to play a viable role in characterizing such a re-brightening of radio emission.
Here, we propose another mechanism that can give rise to the double-peaked radio light curve in core-collapse SNe.
The key ingredient in the present work is to expand the model for the evolution of the synchrotron spectral energy distribution (SED) to a generic form, including fast and slow cooling regimes, as guided by the widely-accepted modeling scheme of gamma-ray burst afterglows. We show that even without introducing an additional CSM shell, the radio light curve would show a double-peaked morphology when the system becomes optically thin to synchrotron self-absorption at the observational frequency during the fast cooling regime.
We can observe this double-peaked feature if the transition from fast cooling to slow cooling regime occurs during the typical observational timescale of SNe.
This situation is realized when the minimum Lorentz factor of injected electrons is initially large enough for the non-thermal electrons' SED to be discrete from the thermal distribution.
We propose SN 2007bg as a special case of double-peaked radio SNe that can be {possibly} explained by the presented scenario.
Our model can serve as a potential diagnostic for electron acceleration properties in SNe.
\end{abstract}


\section{Introduction}
Radio emission from a core-collapse supernova (radio SN) is believed to originate from the interaction between an SN shock and the circumstellar medium (CSM) \citep[e.g.,][]{1982ApJ...259..302C, 1998ApJ...499..810C, 1998ApJ...509..861F, 2006ApJ...641.1029C, 2006ApJ...651..381C, 2012ApJ...758...81M, 2013ApJ...762...14M, 2019ApJ...885...41M, 2020ApJ...898..158M}. The properties of radio SNe depend on the density structure of the CSM, and thus we can use them as an indicator of mass-loss histories of SN progenitors \citep[e.g.,][]{2006ApJ...641.1029C, 2006ApJ...651..381C, 2013MNRAS.428.1207S, 2017ApJ...835..140M, 2021ApJ...918...34M}. Furthermore, modeling radio SNe has a potential to probe the efficiency of the particle acceleration and magnetic field amplification driven by a non-relativistic shock wave \citep{2012ApJ...758...81M, 2014MNRAS.440.2528M, 2019ApJ...874...80M, 2022ApJ...925...48X}. Hence, investigating the nature of radio SNe affords us clues to probing the stellar evolution of massive stars, as well as plasma physics related to shock acceleration mechanism.

\begin{figure}
	\includegraphics[width=\columnwidth]{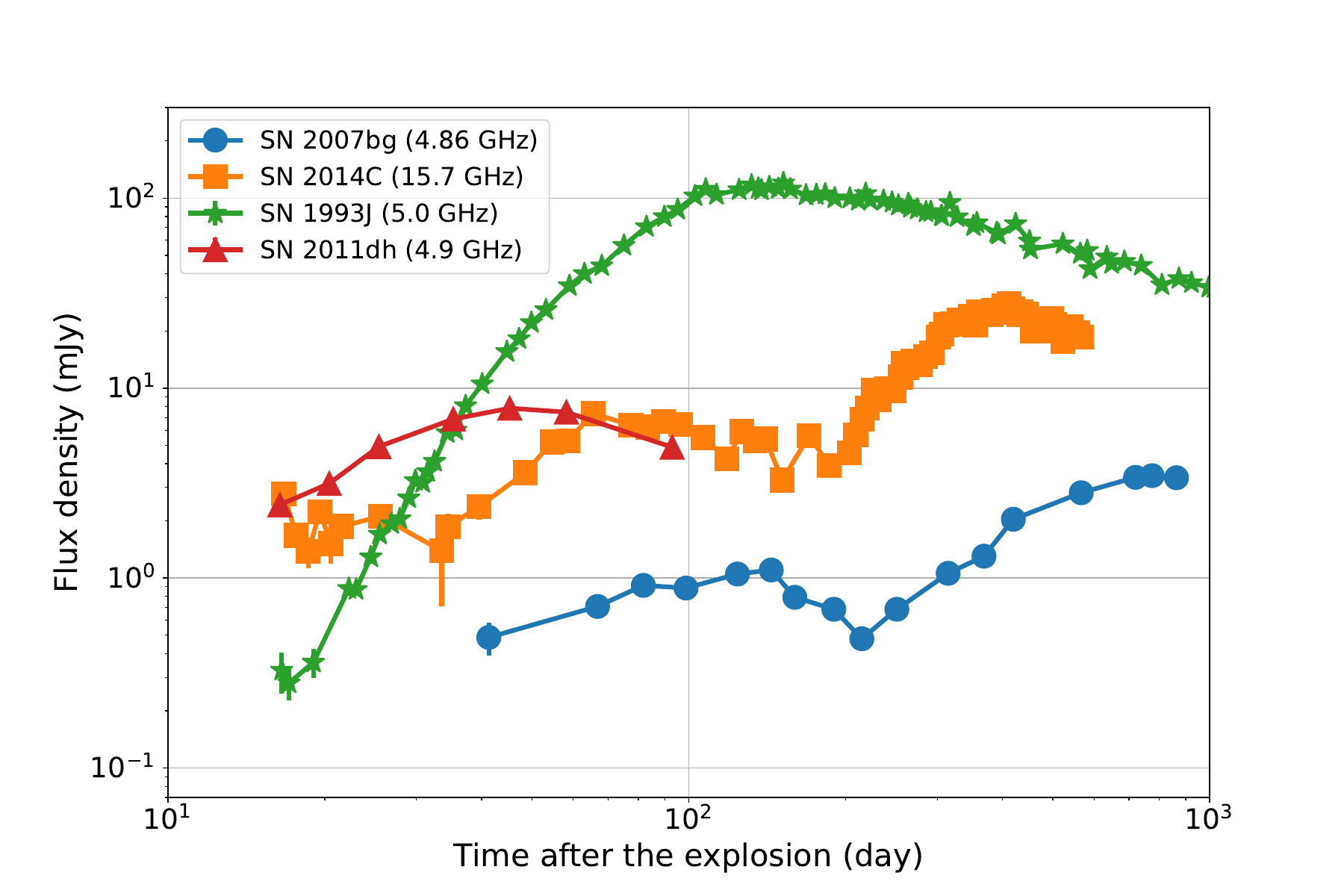}
	\caption{Examples of radio SNe observed {hitherto}. The data are taken from the following references: SN 2007bg from \cite{2013MNRAS.428.1207S}, SN 2014C from \cite{2017MNRAS.466.3648A}, SN 1993J from \cite{2007ApJ...671.1959W}, and SN 2011dh from \cite{2012ApJ...752...78S}.}
	\label{fig:radioSNe_LCs}
\end{figure}

In the classical modeling of radio SNe, the peak luminosity has been determined by the moment when the optical depth to synchrotron self-absorption becomes unity \citep{1998ApJ...499..810C}. Then the light curve is characterized by a single peak.
In fact, this idea has been successful in fitting radio SNe, including SN 1993J and SN 2011dh, on the assumption of (nearly) steady wind CSM configuration \citep[see Figure \ref{fig:radioSNe_LCs} and e.g.,][]{1998ApJ...509..861F, 2012ApJ...752...78S,2012ApJ...758...81M,Chevalier2017}.

On the other hand, there are several objects exhibiting multiple peaks in the radio light curve (see Figure 1).
Examples are broad-lined Type Ic SNe 1998bw and 2007bg \citep{1998Natur.395..663K, 2013MNRAS.428.1207S}. Another object is SN 2014C exhibiting the re-brightening of the radio emission a few years after the explosion (see also \citealt{2009CBET.1714....1S}, \citealt{2018MNRAS.478.5050M}, \citealt{2021ApJ...923...32B} for SN 2004dk, and \citealt{2023ApJ...945L...3M} for SN 2018ivc). Furthermore, SN 2001em and SN 2017ens show luminous radio emissions in the late phase, albeit the detections of radio emissions in the early phase were not reported \citep{2020ApJ...902...55C, 2023arXiv230613730M}.
The boosting of the ejecta velocity has been proposed for SN 1998bw \citep{1999ApJ...526..716L}, while the multiple CSM components including a shell-like structure are preferred for the other objects (see the above references).
We note that the existence of a shell-like CSM component can be an intriguing tracer of the eruptive mass loss of SN progenitors \citep{2017ApJ...835..140M}.

In this study, we present another scenario to reproduce double-peaked radio light curves without introducing complicated setups on SN ejecta and CSM.
The critical point is to take into consideration all of the possible shapes of the synchrotron spectra. The synchrotron spectrum can change its own shape depending on whether the electron spectrum is in the so-called ``fast cooling" or ``slow cooling" regimes, as well as efficiencies of particle acceleration and magnetic field amplification.
This idea has been already introduced in the modeling of gamma-ray burst (GRB) afterglows \citep{1998ApJ...497L..17S, 2008FrPhC...3..306F, 2010ApJ...722..235V, 2020ApJ...896..166R, 2021MNRAS.504.5647S}.
We have applied the same framework to the modeling of radio SNe, incorporating synchrotron self-absorption into the calculation.

We show that a double-peaked radio light curve appears when the system becomes optically thin to synchrotron self-absorption at the observational frequency during the fast cooling regime. The first peak is characterized by the unity of the optical depth during the fast cooling regime, while the second peak appears when the synchrotron frequency related to the minimum Lorentz factor of electrons decreases down to observational frequency during the slow cooling regime.
The observability of the double peak depends on the timescale of the transition from the fast cooling regime to the slow cooling regime. If the transition happens within a typical observational timescale of SNe, current observational instruments can potentially seize both peaks.
This is fulfilled when the spectral energy distribution of non-thermal electrons is separated from the thermal distribution. This originates from the situation where (1) the SN shock transfers a large fraction of its kinetic energy into relativistic electrons, and (2) only a small population of electrons gets involved in shock acceleration.
Our framework can thus provide a new clue to understanding the properties of shock acceleration physics in SNe.

\begin{figure*}
	\includegraphics[width=2\columnwidth]{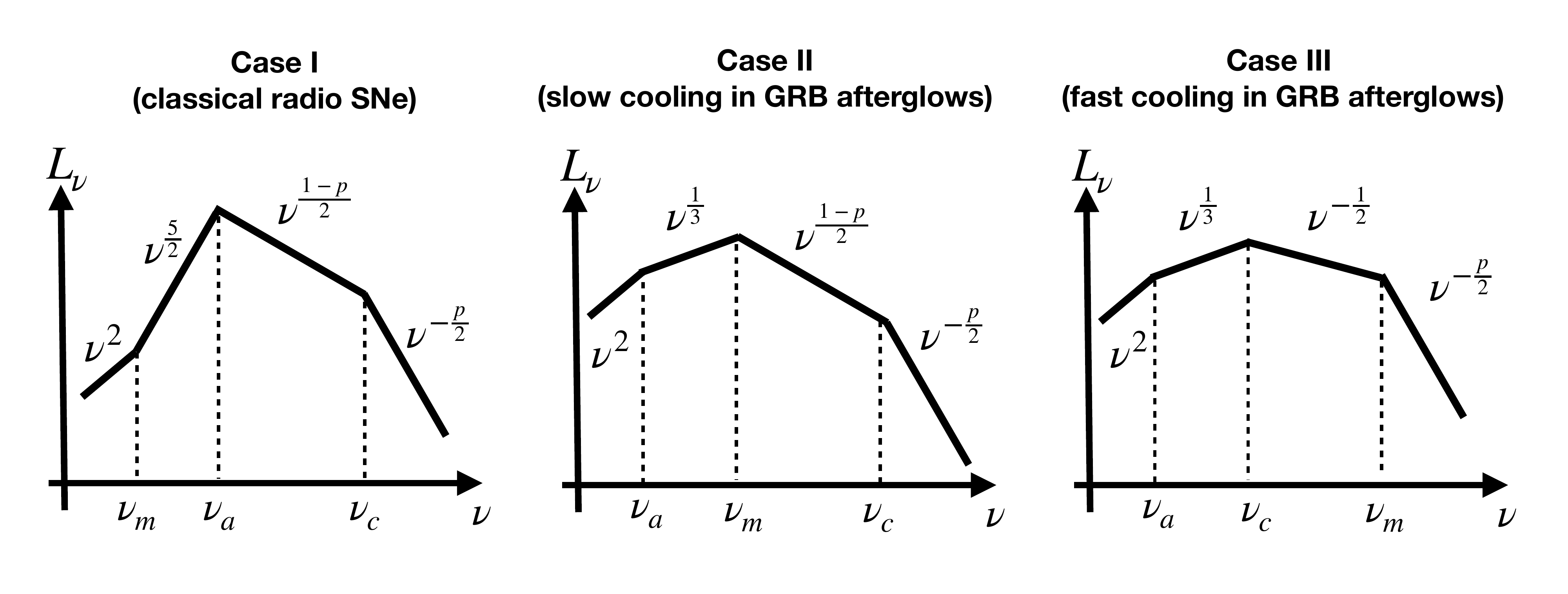}
	\caption{Schematic pictures illustrating the morphologies of the synchrotron spectrum expected in the case of $\nu_a < \nu_c$.}
	\label{fig:concept}
\end{figure*}

This paper is organized as follows. Section \ref{sec:method} illustrates the framework and method of our model, explaining how to compute the luminosity of synchrotron emission from core-collapse SNe.
Section \ref{sec:demonstration} demonstrates our calculation, showing that radio light curves can take two types of morphology depending on electron acceleration efficiencies; a single-peaked and a double-peaked light curve.
The interpretation of each solution is also described.
Section \ref{sec:conditions} depicts the condition for the emergence and the observability of the double peaks in the radio light curve.
Section \ref{sec:implications} discusses implications for astrophysical phenomena and for shock acceleration physics.
Finally, Section \ref{sec:summary} summarizes this paper.

\section{Method}\label{sec:method}

\subsection{Concept}\label{subsec:concept}

This study places emphasis on the evaluation of two characteristic Lorentz factors in a non-thermal electron's spectral energy distribution (SED). One is $\gamma_m$ defined as the minimum Lorentz factor of electrons injected by the SN shock. The other is $\gamma_c$ denoting the cooling break above which synchrotron (or inverse Compton) cooling is faster than adiabatic cooling. Depending on the inequality of $\gamma_m$ and $\gamma_c$, an electron SED can change the shape of itself. An electron SED with $\gamma_m < \gamma_c$ is referred to as ``slow cooling" regime, and that with $\gamma_m > \gamma_c$ as ``fast cooling" regime \citep{1998ApJ...497L..17S}. The regime of the electron SED can indeed affect the resultant synchrotron spectrum \citep[e.g., see][]{2002ApJ...568..820G}. We describe the definitions of $\gamma_m$ and $\gamma_c$, and the function of the electron SED for each regime in Section \ref{subsubsec:gamma_SED}.

Figure \ref{fig:concept} illustrates the schematic pictures of synchrotron spectra treated in this study. In a synchrotron spectrum, the synchrotron frequencies corresponding to $\gamma_m$ and $\gamma_c$ serve as the bending points. We define these frequencies as $\nu_m$ and $\nu_c$. Additionally, the frequency below which optical depth to synchrotron self-absorption exceeds unity characterizes the shape of the spectrum, which we denote as $\nu_a$. In total, three characteristic frequencies become important in a synchrotron spectrum.

Previous studies on radio SNe have implicitly assumed that the peak luminosity is determined by the frequency $\nu_a$ (Case I). The synchrotron spectrum then follows the distribution proportional to $\nu^{5/2}$ and $\nu^{(1-p)/2}$ in the lower and higher frequency ranges than $\nu_a$, respectively, where $p$ is the spectral index of the electron SED at the injection \citep[e.g.,][]{1979rpa..book.....R, 1998ApJ...499..810C, 1998ApJ...509..861F, 2006ApJ...641.1029C, 2006ApJ...651..381C}.
Furthermore, the spectrum in the higher {frequency} range than $\nu_c$ follows the softened distribution {proportional to} $\nu^{-p/2}$.
This method has been used to interpret the observational data of radio SNe \citep[e.g.,][]{2006ApJ...651.1005S, 2012ApJ...758...81M, 2013ApJ...762...14M, 2019ApJ...883..147T}.

However, it should be remarked that this framework requires the assumption that $\nu_m$ must be smaller than $\nu_a$.
\footnote{
The synchrotron spectrum in the frequency range lower than $\nu_m$ follows the distribution proportional to $\nu^2$ \citep{1979rpa..book.....R}, although it has not been taken into consideration in the modeling of radio SNe.
}
If the inequality among these three frequencies changes into $\nu_a < \nu_m < \nu_c$, then the peak luminosity would be characterized by $\nu_m$ and the synchrotron SED in the frequency range $\nu_a < \nu < \nu_m$ becomes proportional to $\nu^{1/3}$ (Case II).
Furthermore when the inequality $\nu_a < \nu_c < \nu_m$ is realized, then the peak luminosity is characterized by the cooling break frequency $\nu_c$ and the spectrum in the frequency range $\nu_c < \nu < \nu_m$ {should be} proportional to $\nu^{-1/2}$ (Case III).
In the modeling of GRB afterglows, the spectra in Case II and Case III are also referred to as ``slow cooling" and "fast cooling`` regime, respectively \citep{1998ApJ...497L..17S}. The radio light curve at a given frequency ($\nu_{\rm obs}$) is determined by which synchrotron spectrum is realized, and by the magnitude of $\nu_{\rm obs}$ compared to the three characteristic frequencies.

This study addresses all of the synchrotron spectra shown in Case I, II, and III. The critical difference from previous models for radio SNe is to incorporate the quantitative evaluation of $\gamma_m$ and $\nu_m$ \citep[however, see also][]{2015ApJ...805..164N}. Depending on the inequality among $\nu_m, \nu_a$, and $\nu_c$, the shape of synchrotron spectra and even the behavior of radio light curves can differ from those in previous studies. Hereafter, we describe our procedures to calculate the luminosity of synchrotron emission in SNe on the basis of this concept.
We note that the synchrotron spectra discussed in this section are restricted in the case of $\nu_a < \nu_c$. It is possible that the situation of the inverse relation ($\nu_c < \nu_a$) is realized depending on the input parameters or setup for the hydrodynamical configuration, which we do not delve into the detailed discussion.


\subsection{Methods}\label{subsec:method}
\subsubsection{Hydrodynamics}\label{subsubsec:hydro}
We consider the situation in which synchrotron emission from SNe is generated at the interaction region between the SN shock and the CSM. This framework has been adopted as the standard model for the observed radio SNe \citep[e.g.,][]{1998ApJ...499..810C, 1998ApJ...509..861F, 2006ApJ...641.1029C, 2006ApJ...651..381C}.
We assume that the SN ejecta is undergoing homologous expansion and its density structure consists of the outer part given by the power-law distribution with the velocity coordinate ($v$) and the inner part with the flat density. The density structure of the outer part is described as follows:
\begin{eqnarray}
\rho_{\rm ej} &=& 
\frac{3M_{\rm ej}}{4\pi v_c^3}
\frac{n-3}{n} t^{-3}
\left(\frac{v}{v_c}\right)^{-n},\\
v_c &=& \left(\frac{10E_{\rm ej}}{3M_{\rm ej}} \frac{n-5}{n-3}\right)^{1/2},
\end{eqnarray}
where $E_{\rm ej}$ and $M_{\rm ej}$ are the kinetic energy and the mass of the ejecta, respectively \citep[for the generalized formula, see e.g,][]{1989ApJ...341..867C, 1999ApJ...510..379M, 2013MNRAS.428.1020M}. The index $n$ determines the density gradient of the ejecta and characterizes the propagation of the SN shock.
As for the CSM structure, we consider the power-law distribution of the CSM density with the index $s$ as follows:
\begin{eqnarray}
\rho_{\rm CSM} = \mathcal{D} r^{-s} = 5\times 10^{-19} \tilde{A}_\ast \left(\frac{r}{10^{15}\,{\rm cm}}\right)^{-s}\,{\rm g}\,{\rm cm}^{-3}.
\end{eqnarray}
Note that it is normalized at the CSM lengthscale of $10^{15}\,{\rm cm}$. Thus, when we assume the CSM velocity of $1000\,{\rm km}\,{\rm s}^{-1}$, $\tilde{A}_\ast=1$ represents a mass-loss episode with the rate of $10^{-5}\,M_\odot\,{\rm yr}^{-1}$ occurring $0.3\,{\rm years}$ before the explosion.
For the case of $s=2$, $\rho_{\rm CSM} = 5\times 10^{11} \tilde{A}_\ast r^{-2}$ is derived and then $\tilde{A}_\ast$ becomes equivalent with the notation $A_\ast$ used in the literature \citep{2012ApJ...758...81M, 2018MNRAS.478..110S}.

Given the density profiles of the ejecta and the CSM, we can write down the time evolution of the SN shock. We employ the power-law solution derived in \cite{1982ApJ...259..302C, 1982ApJ...258..790C} as follows:
\begin{eqnarray}
R_{\rm sh} &=& \left[\frac{(3-s)(4-s)}{(n-3)(n-4)}\frac{u_c^n}{\mathcal{D}}\right]^{\frac{1}{n-s}}
t^{m}, \\
V_{\rm sh} &=& \frac{dR_{\rm sh}}{dt} = \frac{n-3}{n-s}\frac{R_{\rm sh}}{t},\label{3eq:Vsh}
\end{eqnarray}
where $m$ is a deceleration parameter given as $m=(n-3)/(n-s)$. $u_c$ is defined as follows:
\begin{eqnarray}\label{3eq:Ucn}
u_c = 
\left(
\frac{3 M_{\rm ej}}{4\pi v_c^{3-n}}
\frac{n-3}{n}
\right)^{1/n}
.
\end{eqnarray}

\subsubsection{Particle acceleration and magnetic field amplification}\label{subsubsec:PAMFA}

It is believed that an SN shock drives acceleration of charged particles and amplification of turbulent magnetic field during the propagation in the CSM \citep[][]{1978MNRAS.182..147B, 1978MNRAS.182..443B, 1978ApJ...221L..29B, 1983RPPh...46..973D}.
We follow the widely-accepted parameterization \citep[e.g.,][]{Chevalier2017}, in which a fraction of the kinetic energy density dissipated at the SN shock is transported into non-thermal electrons and magnetic field.
We define the energy densities of non-thermal electrons $(u_e)$ and of magnetic field $(u_B)$ as follows:
\begin{eqnarray}
u_e &=& \epsilon_e \rho_{\rm sh} V_{\rm sh}^2, \label{eq:ue}\\
u_B &=& \frac{B^2}{8\pi} = \epsilon_B \rho_{\rm sh} V_{\rm sh}^2, \label{eq:uB}
\end{eqnarray}
where $\rho_{\rm sh} = 9/8 \times \rho_{\rm CSM}(r=R_{\rm sh})$ is the shocked CSM density at the shock radius.\footnote{For the choice of the coefficient in $\rho_{\rm sh}$, there are variations in the definition depending on the papers. See the appendix in \cite{2022ApJ...938...84D}. We follow the parameterization adopted in \cite{2016MNRAS.460...44P}, where the relative velocity of the SN shock to the ambient CSM is taken into account.}
Here we have introduced the parameters $\epsilon_e$ and $\epsilon_B$ that specify the energy transportation efficiencies.
We note that there is still debate regarding the feasible values of these microphysics parameters \citep{2015ApJ...798L..28C, 2015Sci...347..974M,2017PhRvL.119j5101M, 2019MNRAS.485.5105C}.

\subsubsection{Characteristic Lorentz factors and an SED of non-thermal electrons}\label{subsubsec:gamma_SED}

As mentioned in Section \ref{subsec:concept}, an SED of non-thermal electrons is mainly characterized by two Lorentz factors; $\gamma_m$ and $\gamma_c$. {In the phenomenological perspective,} $\gamma_{m}$ {can be} estimated as follows \citep[e.g.,][]{1998ApJ...497L..17S, 2020ApJ...896..166R}:
\begin{eqnarray}\label{eq:gamma_m}
\gamma_m = \left(\frac{p-2}{p-1}\right) \frac{\epsilon_e}{f_e} \frac{\mu_e m_p V_{\rm sh}^2}{m_e c^2},
\end{eqnarray}
where $m_e, m_p, c$ and $\mu_e$ are electron mass, proton mass, light velocity, and the molecular weight of thermal electrons in the upstream, respectively.
We have introduced a parameter $f_e$, the number fraction of non-thermal electrons compared to thermal electrons in the shocked CSM. On the other hand, the cooling break Lorentz factor $\gamma_c$ is determined by comparing the timescale contributed from synchrotron cooling ($t_{\rm syn}$) and inverse Compton cooling ($t_{\rm IC}$) with the adiabatic cooling timescale ($t_{\rm ad}$). Namely, $\gamma_c$ is given as a solution to the equation of
\begin{eqnarray}
\frac{t_{\rm syn}t_{\rm IC}}{t_{\rm syn} + t_{\rm IC}} = t_{\rm ad}.
\end{eqnarray}
If we ignore the contribution from inverse Compton cooling, $\gamma_c$ can be explicitly described as follows:
\begin{eqnarray}
\gamma_c = \frac{6\pi m_e c m}{\sigma_T B^2 t},
\end{eqnarray}
where $\sigma_T$ is the cross section of Thomson scattering.

{We note that the upper bound of the electron SED is characterized by the maximum Lorentz factor of accelerated electrons, which we denote as $\gamma_M$. It is basically determined by the balance between acceleration timescale and synchrotron cooling timescale of electrons \citep[e.g.,][]{1983RPPh...46..973D,2020ApJ...904..188K}. We confirm that the typical SN parameters (e.g., $V_{\rm sh}\sim 10^9\,{\rm cm}\,{\rm s}^{-1}, B\sim 1\,{\rm G}$) lead to $\gamma_M\sim 10^6$, which is sufficiently larger than $\gamma_m$ defined in Equation (\ref{eq:gamma_m}).}

Non-thermal electrons attaining relativistic energies follow a single power-law distribution at the moment of the injection.
We define the SED of relativistic electrons injected per unit time as $C\gamma^{-p}$. The SED is further modified by cooling processes, depending on which $\gamma_m$ or $\gamma_c$ is larger \citep{1998ApJ...497L..17S, 2008FrPhC...3..306F}.
In case of the fast cooling, $\gamma_m > \gamma_c$ is realized and the electron SED is described as
\begin{eqnarray}\label{eq:SEDfast}
N(\gamma) \simeq 
\left\{
\begin{array}{cc}
 {\displaystyle \frac{Ct_{\rm syn}(\gamma=1)\gamma_m^{1-p}}{p-1} \gamma^{-2}} & (\gamma_c < \gamma < \gamma_m)  \\
 {\displaystyle\frac{Ct_{\rm syn}(\gamma=1)}{p-1}\gamma^{-p-1}} & (\gamma_m < \gamma).
\end{array}
\right.
\end{eqnarray}
The low-energy component proportional to $\gamma^{-2}$ is attributed to the strong cooling process so that all of the injected electrons cool down to $\gamma_c$ within the dynamical timescale.
On the other hand, in case of the slow cooling ($\gamma_m < \gamma_c$), the SED can be given as follows:
\begin{eqnarray}\label{eq:SEDslow}
N(\gamma) \simeq 
\left\{
\begin{array}{cc}
 {\displaystyle \frac{Ct_{\rm ad}}{p-1}\gamma^{-p}} & (\gamma_m < \gamma < \gamma_c)  \\
 {\displaystyle \frac{Ct_{\rm syn}(\gamma=1)}{p-1}\gamma^{-p-1}} & (\gamma_c < \gamma).
\end{array}
\right.
\end{eqnarray}
In this regime, only the SED component more energetic than $\gamma_c m_e c^2$ experiences the softening due to the strong cooling. Combining Equations (\ref{eq:SEDfast}) and (\ref{eq:SEDslow}), the unified formula of the electron SED can be described as follows:
\begin{eqnarray}
N(\gamma) = 
\left\{
\begin{array}{cc}
{\displaystyle \frac{C t_{\rm cool}(\gamma)}{(p-1) \gamma}(\max(\gamma, \gamma_m))^{1-p}  }& (\gamma > \min(\gamma_m, \gamma_c))     \\
{\displaystyle 0}  & (\gamma < \min(\gamma_m, \gamma_c))
\end{array}
\right., \nonumber \\
\end{eqnarray}
where $t_{\rm cool}$ is the total cooling timescale of electrons. The abovementioned descriptions on electron SEDs have been often employed in the modeling of GRB afterglows \citep[e.g.,][]{2002ApJ...568..820G, 2010ApJ...722..235V, 2020ApJ...896..166R}, including the synchrotron cooling. In radio SNe, two additional cooling processes can be important. One is inverse Compton scattering through which photons emitted from an SN photosphere would drain energies of relativistic electrons \citep{1998ApJ...509..861F, 2006ApJ...641.1029C, 2006ApJ...651..381C, 2012ApJ...758...81M}. The other is Coulomb interaction that promotes the energy exchange to thermal electrons in a dense circumstellar environment \citep{1998ApJ...509..861F, 2022ApJ...936...98B}.
Then the total cooling timescale is given by the reciprocal sum of each timescale as follows:
\begin{eqnarray}
t_{\rm cool} = \left(\frac{1}{t_{\rm syn}} + \frac{1}{t_{\rm IC}} + \frac{1}{t_{\rm ad}} + \frac{1}{t_{\rm coulomb}}\right)^{-1},
\end{eqnarray}
and the four characteristic timescales are defined as follows:
\begin{eqnarray}
t_{\rm syn} &=& \frac{6\pi m_e c}{\sigma_T}\gamma^{-1}B^{-2}\\
t_{\rm IC} &=& \frac{3\pi m_e c^2}{\sigma_T}\gamma^{-1}L_{\rm bol}^{-1}R_{\rm sh}^2\\
t_{\rm ad} &=& \frac{R_{\rm sh}}{V_{\rm sh}} = \frac{t}{m}\\
t_{\rm coulomb} &=& \frac{m_p m_e^2 c^3}{4\pi e^4\ln\Lambda}\gamma\rho_{\rm sh}^{-1},\label{eq:timescales}
\end{eqnarray}
where $e, L_{\rm bol}$, and $\ln\Lambda=30$ are the elementary charge, bolometric luminosity of the SN, and the coulomb logarithm, respectively \citep[see also][]{2019ApJ...885...41M}. We ignore Klein-Nishina effect because the Lorentz factor of electrons responsible for radio SNe would be $\sim 1000$ (see Section \ref{sec:demonstration}) and the energy of the seed photons is $\sim \mathcal{O}(1\,{\rm eV}$).

\begin{figure*}
    \centering
	\includegraphics[width=1.8\columnwidth]{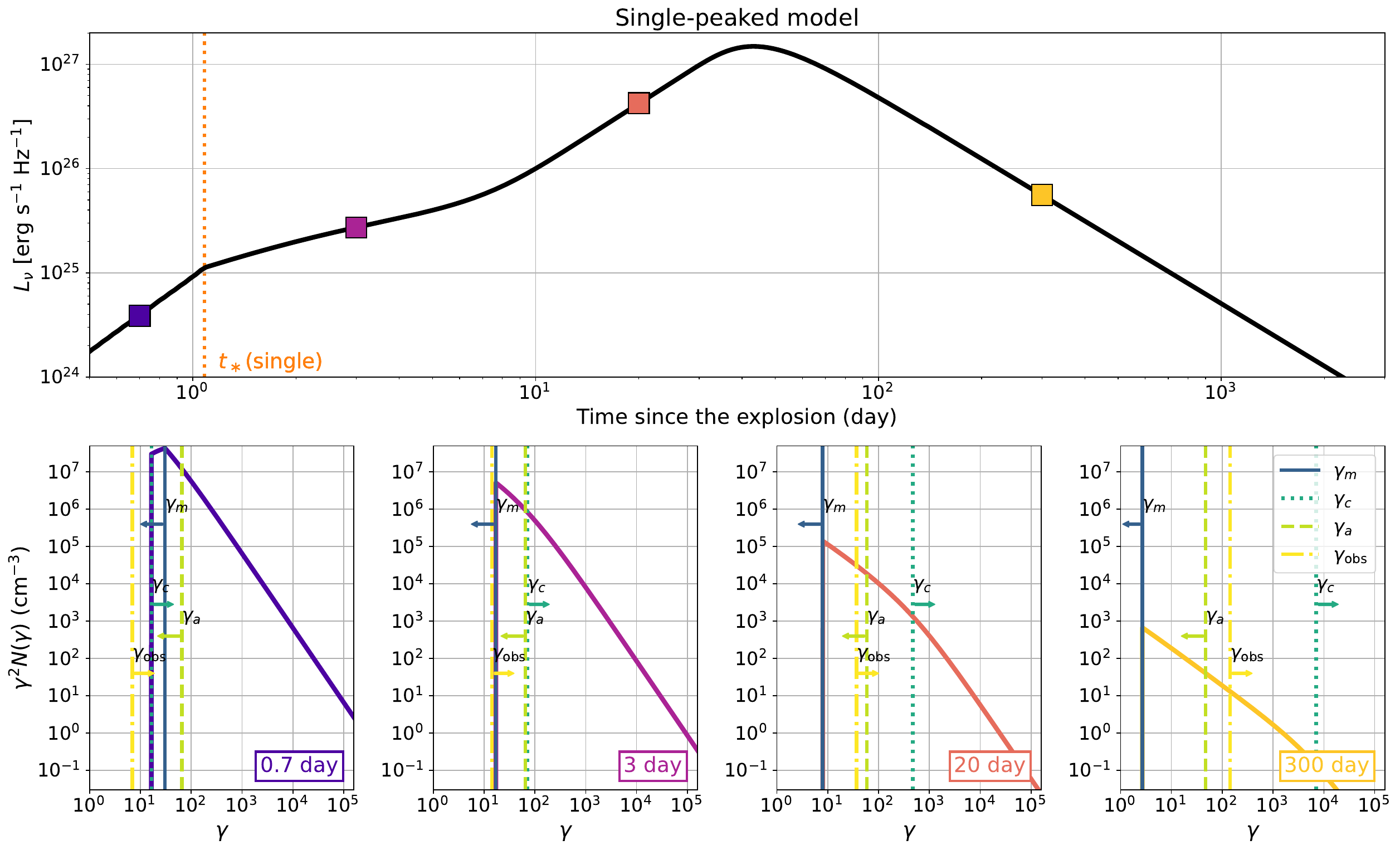}
	\caption{In the top panel, the synthesized radio light curve at the frequency of $5\,{\rm GHz}$ for the model with $(\epsilon_e, f_e) = (10^{-2}, 10^{-2})$ is illustrated. The square symbols denote the moment at which the electron SEDs are inspected in the bottom panels. The dotted vertical line in the light curve panel denotes the regime transition timescale $t_\ast$, which is discussed in Section \ref{sec:conditions} in detail. The vertical lines in the electron SED panels stand for characteristic Lorentz factors of electrons; $\gamma_m, \gamma_c, \gamma_a$, {and $\gamma_{\rm obs}$. Here, $\gamma_a$ is defined as }the Lorentz factor whose corresponding synchrotron frequency characterizes the unity of the optical depth to synchrotron self-absorption.
 } 
	\label{fig:single_LCNgamma}
\end{figure*}

\begin{figure*}
    \centering
	\includegraphics[width=1.8\columnwidth]{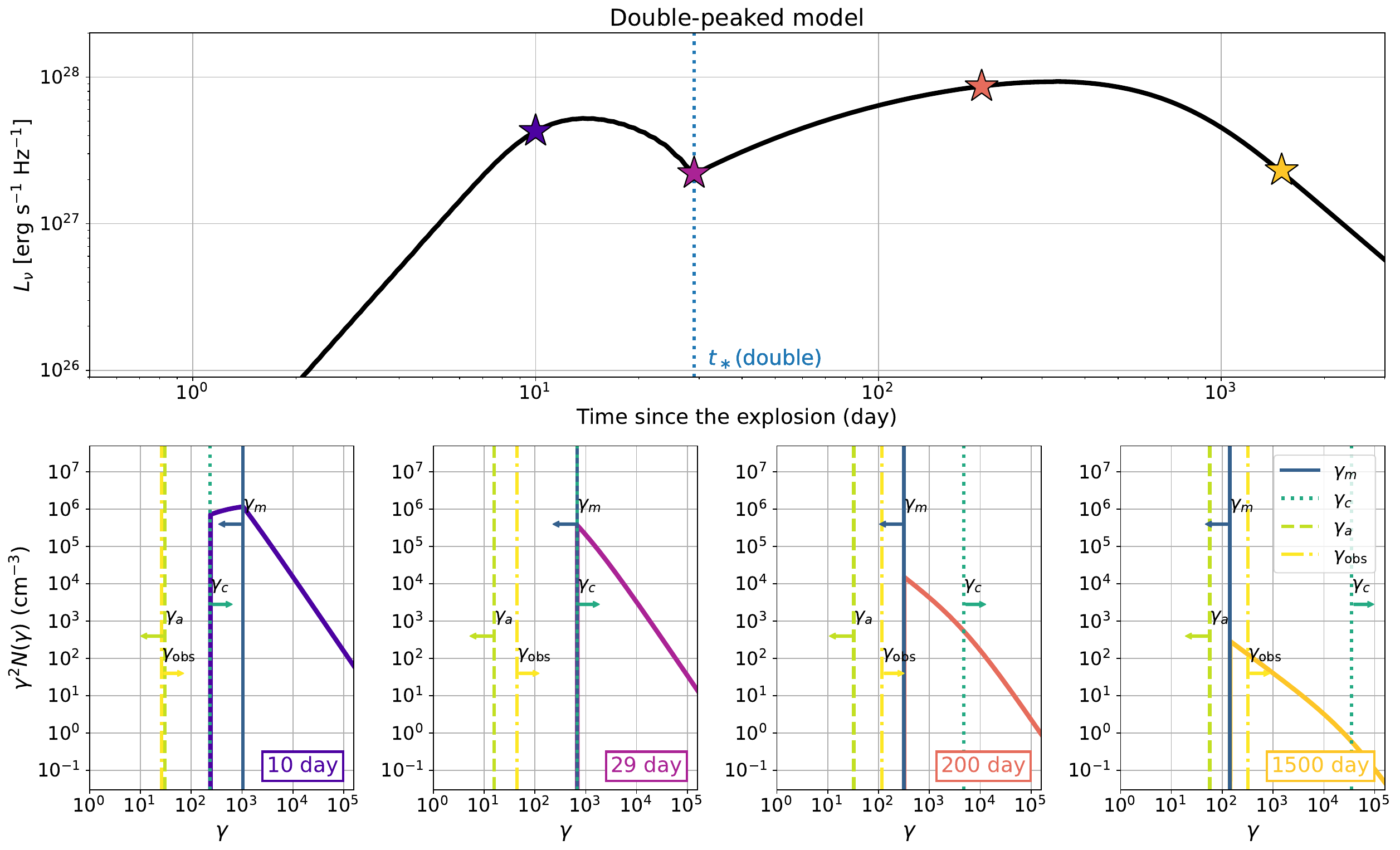}
	\caption{Same as Figure \ref{fig:single_LCNgamma}, but for the double-peaked model with $(\epsilon_e, f_e) = (10^{-1}, 10^{-3})$. The star symbols are used instead of the square symbols, to point out the moment when the electron SEDs are displayed in the bottom panel.}
	\label{fig:double_LCNgamma}
\end{figure*}

Finally, we give the definition of $C$ as follows:
\begin{eqnarray}
C = \frac{(p-2)u_e}{\gamma_m^{2-p}m_e c^2}\frac{V_{\rm sh}}{\Delta R_{\rm sh}},
\end{eqnarray}
where $\Delta R_{\rm sh}$ is the geometrical thickness of the shocked CSM region.
We assume $\Delta R_{\rm sh}$ is proportional to $R_{\rm sh}$, where the coefficient refers to the results in previous studies \citep[][]{1982ApJ...259..302C, 1994ApJ...420..268C, 1999ApJS..120..299T}. Typically it falls within the range of $\sim 0.2-0.5$ depending on the density gradient of the ejecta $(n)$.

We note that when the CSM follows the steady wind configuration, there must be a phase when $\gamma_m$ is larger than $\gamma_c$, with the electron SED following the fast cooling regime, toward $t\rightarrow 0$. This is because $\gamma_m$ decreases {with time in our phenomenological definition} while $\gamma_c$ increases as a function of time. We refer readers to Appendix \ref{app:proportional}, which specifies the proportional dependencies of the quantities including $\gamma_m$ and $\gamma_c$.

\subsubsection{Synchrotron emission}\label{subsubsec:synchrotron}
Given the strength of the magnetic field and the profile of the non-thermal electrons' SED in the shocked CSM region, we can estimate the quantities related to synchrotron emission \citep[e.g.,][]{1979rpa..book.....R}. The synchrotron spectrum emitted by one electron $(P_{\rm syn}(\gamma))$ is described as a function of the observational frequency $\nu$,
\begin{eqnarray}
P_{\rm syn}(\gamma) = \frac{\sqrt{3}e^3B}{m_e c^2}G\left(\frac{\nu}{\nu_{\rm syn}(\gamma)}\right),\label{eq:Psyn}
\end{eqnarray}
where the following quantities are defined:
\begin{eqnarray}
\nu_{\rm syn}(\gamma) &=& \frac{3\gamma^2 eB}{4\pi m_e c}, \\
G(x) &=& \frac{1}{2}\int_0^x F\left(\frac{x}{\sin\theta}\right)\sin^2\theta d\theta, \\
F(x) &=& x \int_x^\infty K_{5/3}(y) dy.
\end{eqnarray}
$K_{5/3}(y)$ is the modified Bessel function with the order of 5/3.
Here, the dependence on the pitch angle ($\theta$) of emitter electrons \citep{1979rpa..book.....R} is absorbed into $G(x)$, by taking the average over the pitch angle. We employ the fitting formula for $G(x)$ suggested in \cite{2010PhRvD..82d3002A}.

The emissivity $(j_{\rm syn})$ and the absorption coefficient to synchrotron self-absorption $(\alpha_{\rm syn})$ are calculated as follows \citep{1979rpa..book.....R}:
\begin{eqnarray}
j_{\rm syn} &=& \frac{1}{4\pi} \int P_{\rm syn}(\gamma) N(\gamma) d\gamma,\label{eq:jsyn} \\
\alpha_{\rm syn} &=& \frac{-1}{8\pi m_e \nu^2} \int P_{\rm syn}(\gamma) \gamma^2 \frac{\partial}{\partial \gamma}\left(\frac{N(\gamma)}{\gamma^2}\right) d\gamma.
\end{eqnarray}
Then the optical depth to synchrotron self-absorption can be determined by
\begin{eqnarray}
\tau_{\rm syn} = \alpha_{\rm syn} \Delta R_{\rm sh}.
\end{eqnarray}
Finally, the luminosity is calculated as follows:
\begin{eqnarray}
L_{\nu} = 4\pi^2 R_{\rm sh}^2 \frac{j_{\rm syn}}{\alpha_{\rm syn}} (1-e^{-\tau_{\rm syn}}).
\end{eqnarray}

\section{A double-peaked radio light curve}\label{sec:demonstration}
In this section, we show examples of our radio light curve models.
For simplicity, we parametrize only $\epsilon_e$ and $f_e$, which are related to the efficiency of electron acceleration.
We fix the other parameters as follows: $M_{\rm ej} = 2\,M_\odot, E_{\rm ej} = 10^{51}\,{\rm erg}, \tilde{A}_\ast=50, \epsilon_B = 0.005, p=3, s=2$, and $n=7$. These values are in line with the typical parameter choice for stripped-envelope SNe \citep{1999ApJ...510..379M, 2006ApJ...651..381C, 2016MNRAS.457..328L, 2019ApJ...883..147T, 2021ApJ...918...34M}. In addition, we neglect the contribution from inverse Compton cooling for simplicity as it depends on the time evolution of the bolometric luminosity of an SN itself.

The upper panels of Figure \ref{fig:single_LCNgamma} and \ref{fig:double_LCNgamma} show the radio light curves at the frequency of $\nu_{\rm obs}=5\,$GHz for the models with the parameter set of $(\epsilon_e, f_e) = (10^{-2}, 10^{-2})$ and $(10^{-1}, 10^{-3})$, respectively.
We can obviously see that the light curves are different from each other, even though the parameters except for $\epsilon_e$ and $f_e$ are the same. Indeed, these two parameters can affect the value of $\gamma_m (\propto \epsilon_e f_e^{-1})$ and the regime of the electron SED. This {yields} a diversity of the radio light curve morphology.

\begin{figure*}[!ht]
    \centering
	\includegraphics[width=2\columnwidth]{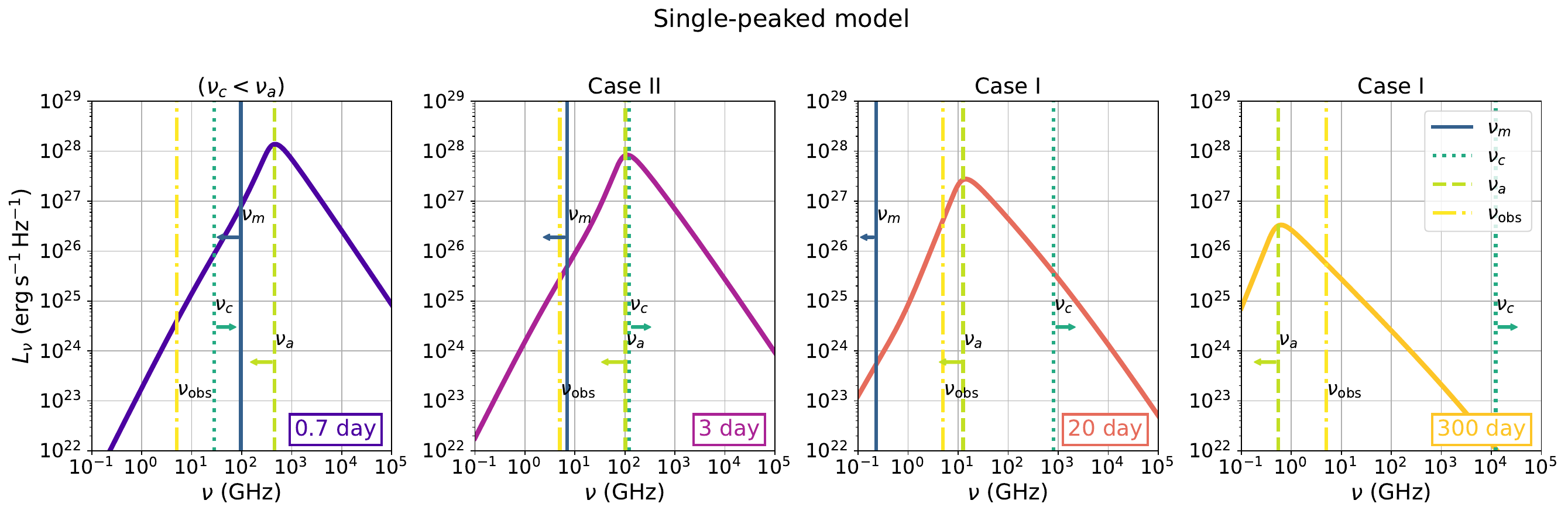}
	\caption{The SEDs of synchrotron emission at several ages are shown for the single-peaked model. The time windows displayed here correspond to the square symbols in Figure \ref{fig:single_LCNgamma}. The vertical lines denote characteristic frequencies; $\nu_m, \nu_c, \nu_a$, and $\nu_{\rm obs}$.
 }
	\label{fig:synchrotronSEDs_single}
\end{figure*}

Comparing the non-thermal electrons' SEDs between the two models allows us to investigate what gives rise to the difference.
The lower panels in Figure \ref{fig:single_LCNgamma} show the SEDs of non-thermal electrons at several ages for the model with $(\epsilon_e, f_e) = (10^{-2}, 10^{-2})$. The corresponding synchrotron SEDs are displayed in Figure \ref{fig:synchrotronSEDs_single}. The electron SED is in the fast cooling regime in the beginning phase ($t = 0.7\,{\rm day}$), but immediately evolves into the slow cooling regime ($t=3\,{\rm days}$). In the early phase, the Lorentz factor responsible for the emission {at the observational frequency} $\gamma_{\rm obs} \simeq (4\pi m_e c \nu_{\rm obs} / 3eB)^{1/2}$ is smaller than $\gamma_m$, and the radio luminosity is mainly covered by electrons with $\gamma \sim \gamma_m$.
Due to the optically thick situation, the synchrotron emission follows the distribution with the relationship of $L_\nu \propto \nu^2$ (Case II).
Once $\gamma_m$ decreases below $\gamma_{\rm obs}$, non-thermal electrons with $\gamma\sim\gamma_{\rm obs}$ become main emitters of synchrotron emission at $5\,{\rm GHz}$ ($t=20\,{\rm days}$).
The system is still optically thick at $\nu=\nu_{\rm obs}$ and $L_\nu$ is proportional to $\nu^{5/2}$ (Case I).
The peak luminosity is characterized by the unity of the optical depth. After the system becomes optically thin, the radio luminosity begins to decay with time ($t=300\,{\rm days}$).
Except for the initial phase, this behavior is compatible to the classical interpretation of radio SNe observed ever \citep[e.g.,][]{2006ApJ...651..381C, 2006ApJ...641.1029C, 2012ApJ...752...78S, 2020ApJ...903..132H}.

\begin{figure*}[!ht]
    \centering
	\includegraphics[width=2\columnwidth]{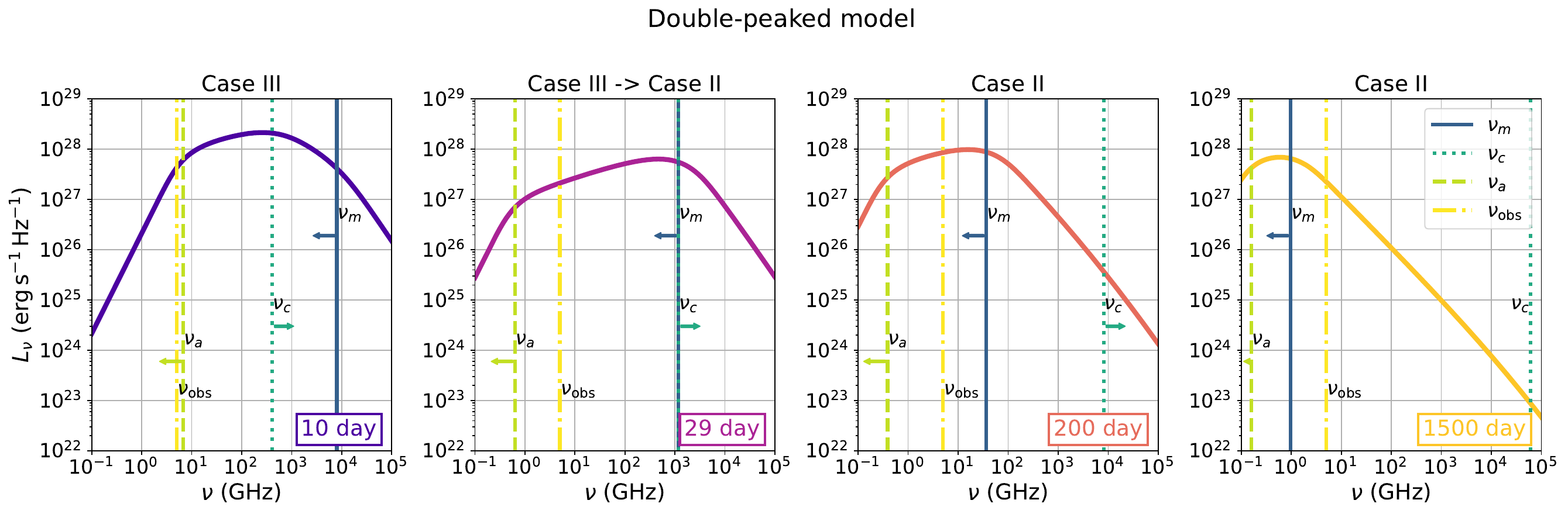}
	\caption{Same as Figure \ref{fig:synchrotronSEDs_single}, but for the double-peaked model. The displayed time windows correspond to the moment pointed out by the star symbols in Figure \ref{fig:double_LCNgamma}.}
	\label{fig:synchrotronSEDs_double}
\end{figure*}

The situation changes in the double-peaked model, where the electron SED remains in the fast cooling regime for a longer timescale as illustrated in the lower panels of Figure \ref{fig:double_LCNgamma}. The resultant synchrotron SEDs are displayed in Figure \ref{fig:synchrotronSEDs_double}. As $\gamma_{\rm obs}$ is smaller than $\gamma_c$\footnote{Note that in this case $\gamma_c < \gamma_m$ is realized.}, electrons with $\gamma\sim\gamma_c$ become main emitters of synchrotron radiation at $\nu=\nu_{\rm obs}$.
The system is optically thick at this frequency and the synchrotron SED takes the form in Case III ($t=10\,{\rm days}$).
However, the density of the CSM swept up by the shock decreases with time. Thus once the system becomes optically thin at $\nu=\nu_{\rm obs}$, the radio emission turns into decay. This is the physical process characterizing the first peak in the double-peaked radio light curve.

Since $\gamma_c$ increases with time due to the weakening of the synchrotron cooling, $\gamma_c$ meets up with $\gamma_m$ soon after the first peak, and this moment characterizes {the dip,} the local minimum of the radio light curve  ($t= 29\,{\rm days}$). After that, the electron spectrum enters the slow cooling regime.
As $\gamma_{\rm obs} < \gamma_m$ is satisfied even immediately after the transition, electrons with $\gamma\sim\gamma_m$ would be the main emitters of synchrotron radiation. The important effect is that $\gamma_m$ decreases with time, resulting in the rapid increase in the population of the emitter electrons. This affords the re-brightening of radio emission ($t= 200\,{\rm days}$). Once $\gamma_{\rm obs} \simeq \gamma_m$ is achieved, the second peak is shaped, and the synchrotron SED evolves into the spectrum in Case II.
After the second peak, the emission is responsible for electrons with $\gamma\sim\gamma_{\rm obs}$, and since the density of the CSM continues to decline with time, the emission results in decay from the second peak ($t= 1500\,{\rm days}$). We refer readers to Appendix \ref{app:derivation} for the mathematical interpretation of the origin of the double peak. {We also mention that the possibly similar double-peaked characteristic has appeared in the numerical modeling of \citet{2023MNRAS.524.6004W}, where multi-wavelength afterglow emission associated with a magnetar giant flare or a fast radio burst is calculated on the basis of synchrotron emission in a similar way to our study.}

We note that the critical difference between the two models lies in the timing of the regime transition. In the single-peaked model, the transition occurs during the optically thick phase, while in the double-peaked model, it happens in the optically thin phase.
This means that all of the peaks discussed in this section are shaped in qualitatively different ways from each other.
In a single-peaked model, the moment of $\tau_{\rm syn}=1$ during the slow-cooling regime is important for the formation of the peak, whereas in a double-peaked model, the moments of $\tau_{\rm syn}=1$ during the fast-cooling regime and of $\gamma_{\rm obs}\simeq \gamma_m$ are critical.
This also indicates that the difference between these two models would emerge even in the synchrotron spectra; Case I is realized in the single-peaked model, while Case II and III would be realized in the double-peaked model.

\begin{figure*}[!ht]
    \centering
	\includegraphics[width=0.666\columnwidth]{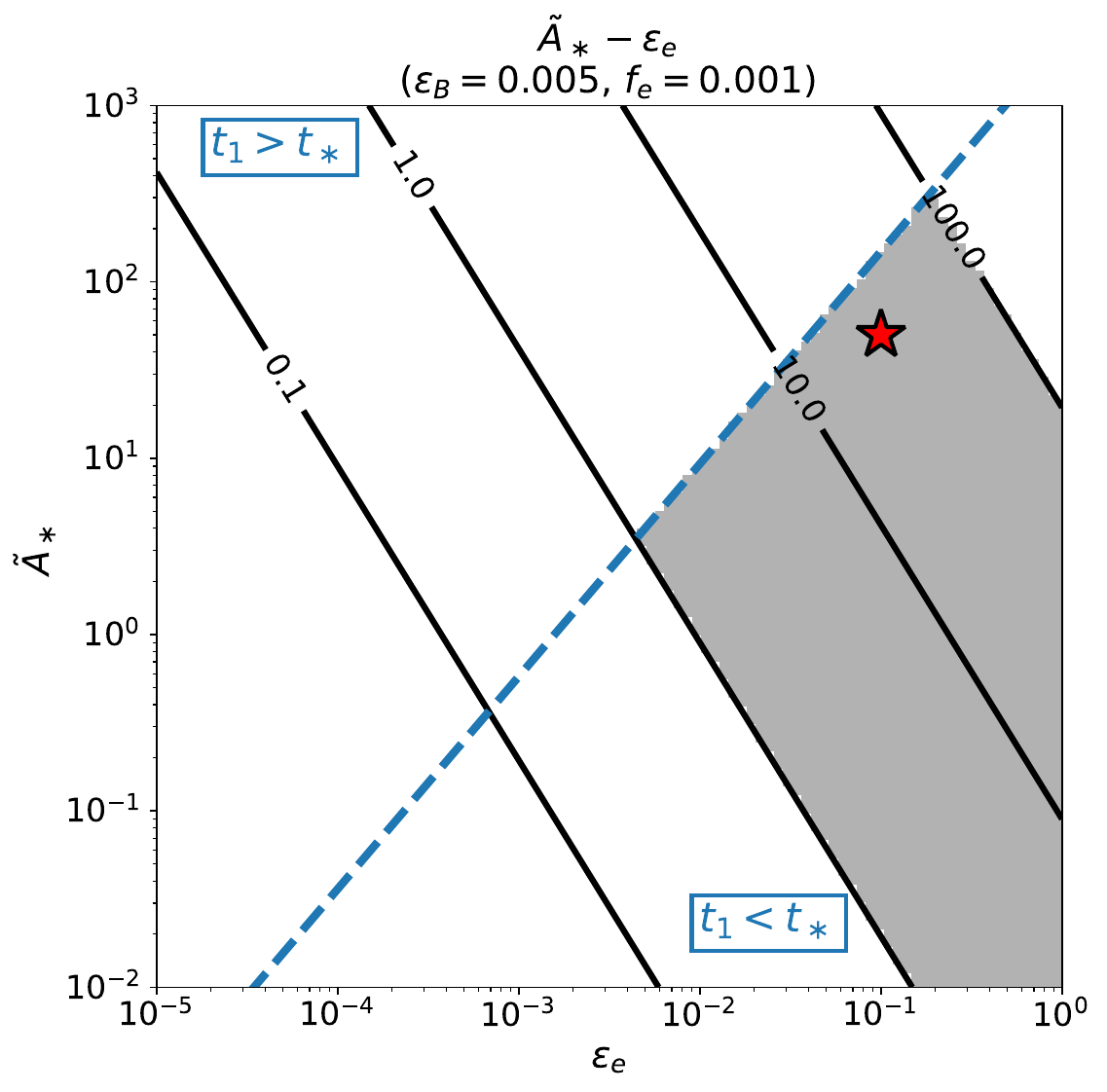}
	\includegraphics[width=0.666\columnwidth]{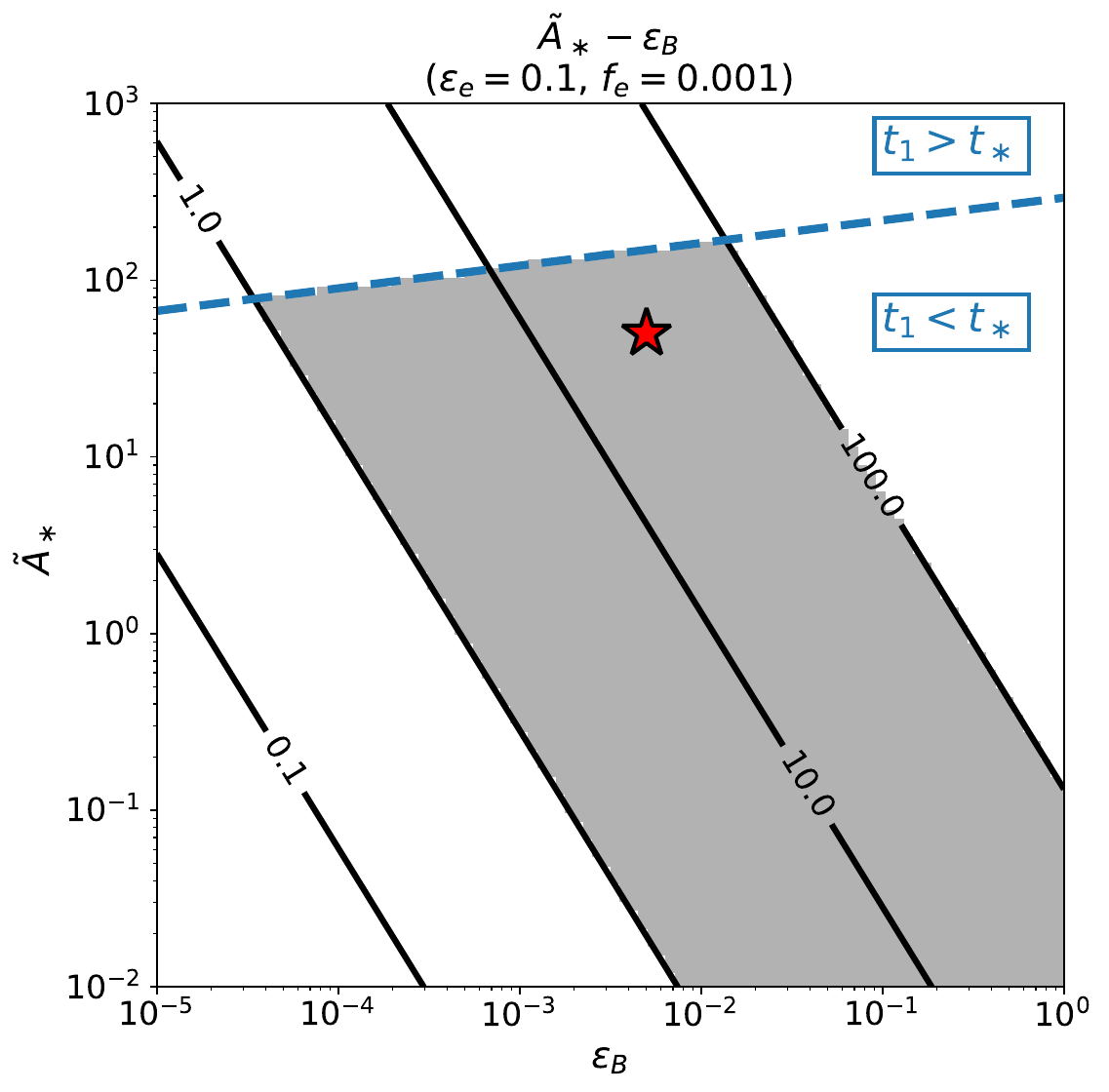}
	\includegraphics[width=0.666\columnwidth]{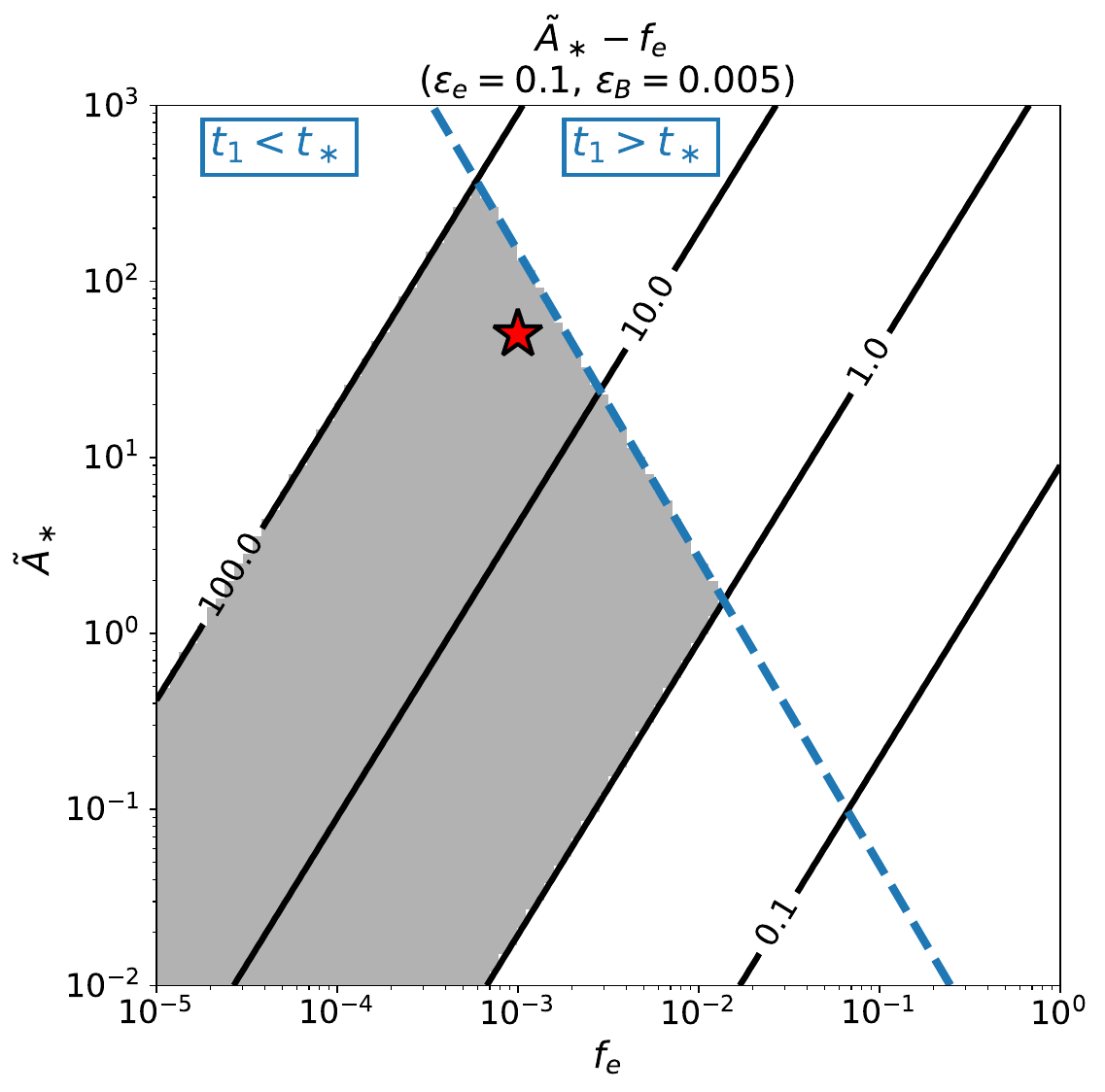}
	\caption{Conditions for the parameters exhibiting {observable} double-peaked light curves at the frequency $\nu_{\rm obs}=5\,{\rm GHz}$. The black lines illustrate the contours of $t_\ast$ in the unit of day, while the blue dashed line exhibits the critical parameter sets given by $t_1 = t_\ast$, which invoke the regime transition and the decline of the optical depth below unity simultaneously. To clearly show the parameter dependence, we draw the parameter variations of $t_\ast$ as functions of $\tilde{A}_\ast$ and $\epsilon_e$ (left), $\epsilon_B$ (middle), and $f_e$ (right), {and the other parameters are fixed (see the main text)}. The regions below the critical line satisfy the condition of $t_1 < t_\ast$; the system becomes optically thin to synchrotron self-absorption at $\nu=\nu_{\rm obs}$ during the fast cooling regime. The parameter spaces $1\,{\rm day} < t_\ast < 100\,{\rm days}$ as a typical observational timescale of radio SNe \citep[e.g.,][]{2021ApJ...908...75B}, with the above condition fulfilled, are exhibited by the gray-shaded regions.
 }
	\label{fig:conditions}
\end{figure*}

\section{Conditions for the emergence of double peaks}\label{sec:conditions}

In the previous section we showed that depending on the particle acceleration efficiencies, radio emission from SNe can reach the peak twice. It is useful to clarify what kinds of parameter sets lead to a double-peaked light curve. For this purpose, we define two characteristic timescales in radio light curves. One is $t_1$ when the system becomes optically thin to synchrotron self-absorption at $\nu=\nu_{\rm obs}$. $t_1$ cannot be explicitly written down, but derived as a solution of the equation $\tau_{\rm syn} = 1$.

Another timescale is $t_\ast$, at which the electron SED transits from fast cooling to slow cooling regime. When we neglect the cooling effect from inverse Compton scattering, $t_\ast$ can be analytically described as follows:
\begin{eqnarray}\label{eq:t_ast}
t_\ast = \left[ \left(\frac{p-2}{p-1}\right) \left(\frac{\mu_e m_p \sigma_T}{6\pi m_e^2 c^3 m}\right) V_0^2 B_0^2 \frac{\epsilon_e \epsilon_B}{f_e} \tilde{A}_\ast^{\frac{n-4}{n-s}} \right]^{\frac{1}{-4m+ms+3}}, \nonumber \\
\end{eqnarray}
where the shock velocity and the magnetic field are rewritten as $V_{\rm sh} = V_0 \tilde{A}_\ast^{\frac{-1}{n-s}} t^m$ and $B = B_0 \epsilon_B^{\frac{1}{2}}{\tilde{A}_\ast}^{\frac{n-2}{2(n-s)}} t^{\frac{m(2-s)}{2}-1}$, respectively (see also Appendix \ref{app:proportional}). $V_0$ and $B_0$ stand for the coefficients of the shock velocity and the magnetic field strength that are connected to the CSM density, time, and $\epsilon_B$. We note that $t_\ast$ is independent of $\nu_{\rm obs}$ unlike $t_1$.

The condition for the emergence of the double-peaked radio light curve is that the system should become optically thin at $\nu=\nu_{\rm obs}$ during the fast cooling regime. Namely, $t_1 < t_\ast$ should be satisfied. {Furthermore, in order to observationally capture the double peak feature, $t_\ast$ needs to be comparable to the typical observational timescale of SNe. In short, $t_\ast \sim t_{\rm obs}$ is required, where $t_{\rm obs}\sim 1-100\,{\rm days}$ is a typical observational timescale of radio SNe \citep{2021ApJ...908...75B}. Otherwise, we would miss the moment of the regime transition.}
These two requirements are important for the emergence of the observable double peak in the radio light curve.

Figure \ref{fig:conditions} illustrates the parameter spaces that result in the observable double-peaked radio light curve at $\nu_{\rm obs}=5\,{\rm GHz}$.
Here we focus on the dependencies of $t_\ast$ on the CSM density and microphysics parameters $\epsilon_e, \epsilon_B$, and $f_e$. The other parameters ($M_{\rm ej}, E_{\rm ej}, p, s$, and $n$) are fixed to those in the double-peaked model in Section \ref{sec:demonstration} to clearly show the parameter dependencies.
We also plot the blue dashed lines to represent the critical parameter sets that cause the regime transition and the drop of the optical depth beneath unity simultaneously.
This is determined as a solution to the equation of $t_1 = t_\ast$.

The gray-shaded regions in Figure \ref{fig:conditions}, where the double-peaked radio light curve can be observed, are obtained by the following conditions.
First, to make the regime transition during the fast cooling we need to go into the parameter space below the critical line ($t_1 < t_\ast$). Second, we require $t_\ast$ to be moderately long so that radio observations can trace the moment of the regime transition.
From Figure \ref{fig:conditions}, we can infer that large $\epsilon_e$ and small $f_e$ are preferred for the emergence of the observable double-peaked light curve.
This parameter choice leads to high $\gamma_m$ and corresponds to the situation where a substantial amount of energy from the shock is deposited into the limited population of non-thermal electrons. We also confirm that the emergence of the double peak prefers a gentle gradient of the SN ejecta (smaller $n$).

\begin{figure}
    \centering
	\includegraphics[width=\columnwidth]{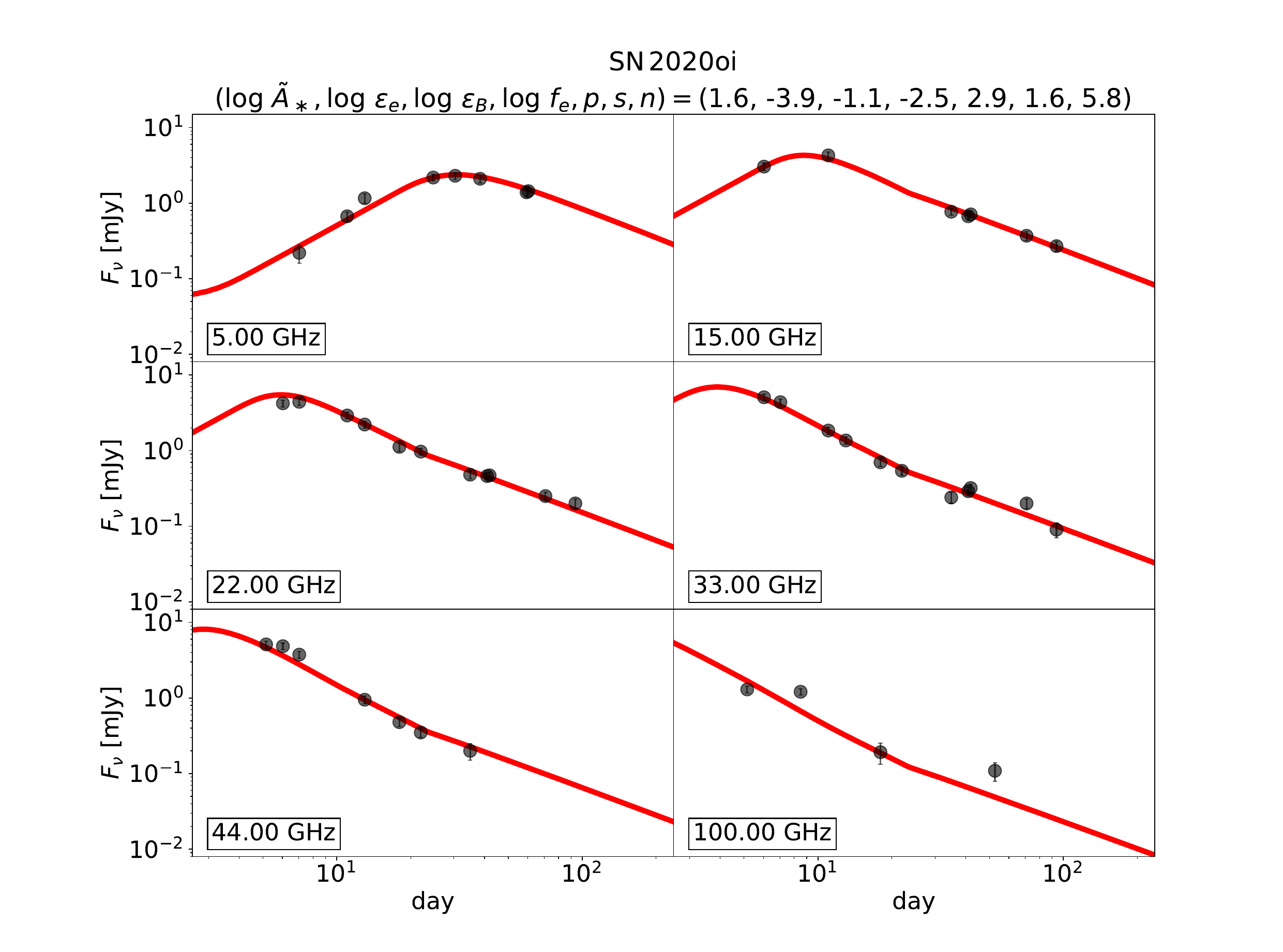}
	\includegraphics[width=\columnwidth]{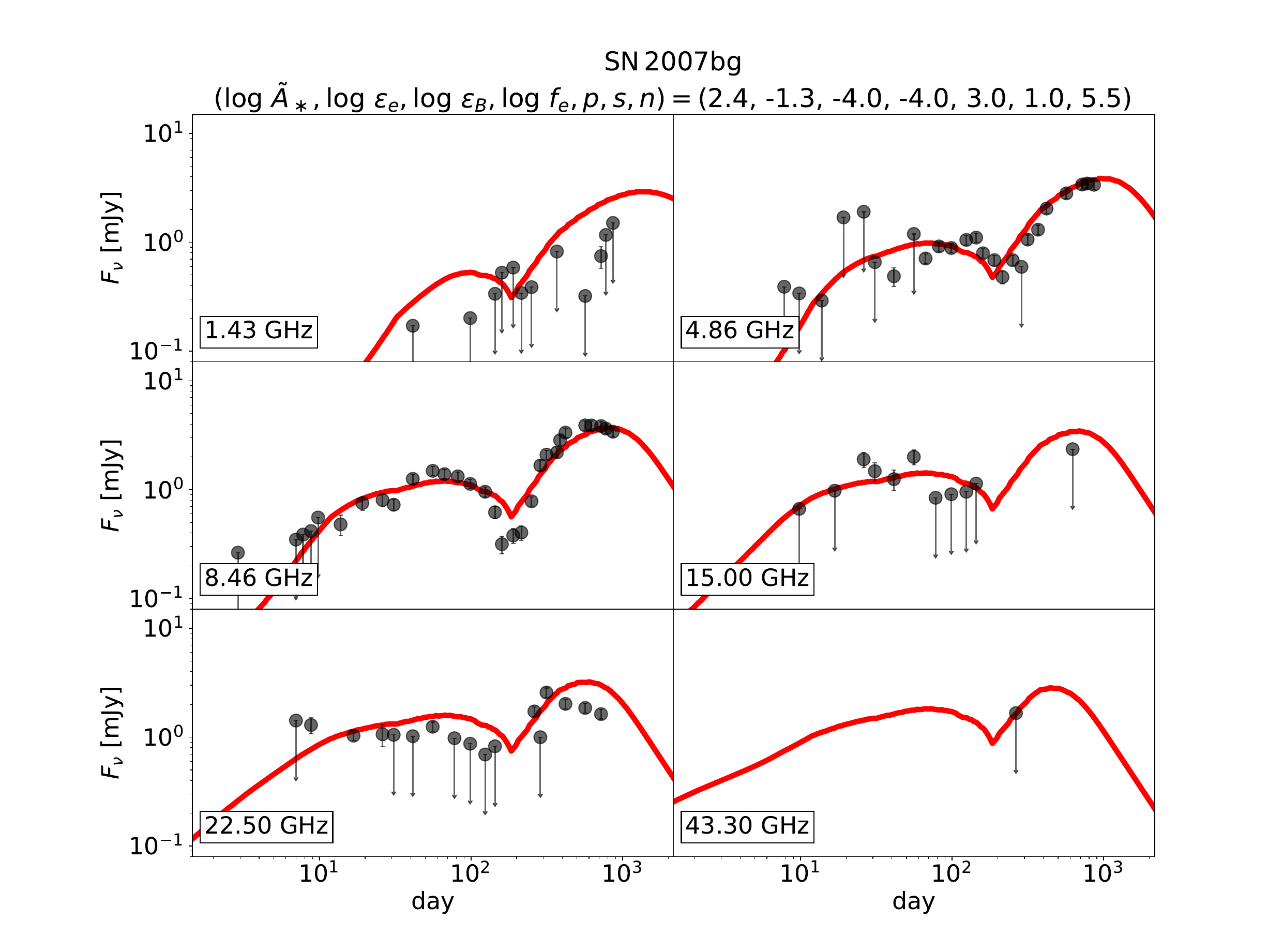}
	\caption{The radio light curves synthesized in our model are compared with observational data of SN 2020oi (top) and SN 2007bg (bottom). The parameter sets used in the calculations are shown in the upper part of each panel. Data points refer to the following studies; \protect\cite{2020ApJ...903..132H, 2021ApJ...918...34M} for SN 2020oi, and \protect\cite{2013MNRAS.428.1207S} for SN 2007bg.}
	\label{fig:fitting}
\end{figure}

\section{Discussion}\label{sec:implications}

\subsection{Reproduction of the radio SN data and implications for observations}\label{subsec:observation}

Radio SNe observed so far have been fitted with the spectrum of Case I in Figure \ref{fig:concept} \citep[e.g.,][]{1998ApJ...499..810C, 1998ApJ...509..861F, 2006ApJ...641.1029C, 2006ApJ...651..381C, 2012ApJ...758...81M, 2013ApJ...762...14M, 2019ApJ...883..147T, 2020ApJ...903..132H}.
In this case, $\nu_m < \nu_a < \nu_c$ is implicitly assumed.
On the other hand, the model presented in this study can be regarded as the application of the modeling of GRB afterglows, where the synchrotron SED is determined depending on the inequality of characteristic synchrotron frequencies \citep[see e.g.,][]{1998ApJ...497L..17S, 2002ApJ...568..820G, 2020ApJ...896..166R}.
This means that our method is applicable to all types of spectra (Case I, II, and III), while we need to check whether our model can reproduce the observational data, as well as the consistency with previous studies.

Figure \ref{fig:fitting} shows the fitting results of the observed radio data of SN 2020oi and SN 2007bg based on our model.
These models take into account the cooling timescale of inverse Compton scattering induced by seed photons from the SN photosphere.
SN 2020oi is a Type Ic SN discovered in M100 \citep{2021ApJ...908..232R, 2022ApJ...924...55G}, and radio observations have been conducted at the frequencies from $5\,{\rm GHz}$ to $100\,{\rm GHz}$ \citep{2020ApJ...903..132H, 2021ApJ...918...34M}.
The observed radio light curve is characterized by a single peak.
SN 2007bg is a broad-lined Type Ic SN \citep{2010A&A...512A..70Y, 2013MNRAS.428.1207S}.
Particularly \cite{2013MNRAS.428.1207S} argued that the radio light curve in SN 2007bg exhibits the prominent double peak at $4.86$ and $8.46\,{\rm GHz}$, and proposed the two components of the CSM around the progenitor as the origin of the multiple peaks.

As seen in Figure \ref{fig:fitting}, our computations can successfully reproduce the models consistent with radio data of both the single-peaked radio SN 2020oi and the double-peaked radio SN 2007bg. We confirm that the synchrotron SED at the peak in SN 2020oi is fitted with the spectrum of Case I, demonstrating consistency with previous studies on radio SNe \citep{2021ApJ...918...34M}. Furthermore, the model for SN 2007bg can reproduce the double-peaked radio light curve without invoking the {multiple components of the CSM}. Instead, the double-peaked feature is formed by the transition of the non-thermal electrons' spectrum from the fast cooling to the slow cooling regime. As expected, the model that fits the light curve of SN 2007bg has a high value of $\epsilon_e$ relative to $f_e$ ($\epsilon_e \simeq 5\times 10^{-2}, f_e\simeq 10^{-4}$). We suggest that {our model} serves as an alternative explanation of the double-peaked radio light curve.

We note that SN 2007bg may be a special case of double-peaked radio SNe fitted by our model. As seen in Appendix \ref{app:derivation}, our model reproduces the luminosity variation only by a factor of a few because the luminosity can be described by the power-law terms. Thus, the regime transition of the electron SED may be capable of explaining double-peaked radio SNe with a luminosity variation within a factor of a few. On the other hand, it is also known that there are some radio SNe exhibiting the re-brightening by orders of magnitude (e.g., \citealt{2017ApJ...835..140M} for SN 2014C, and \citealt{2023ApJ...945L...3M} for SN 2018ivc). We suppose that these SNe are beyond the explanation by our model, and then other physical processes may be important; the inhomogeneous CSM density structure \citep{2021ApJ...918...34M, 2023ApJ...942...17M} or the time-dependent microphysical parameters \citep{2023arXiv230208230H} would be feasible candidates.

We suggest the multi-wavelength observations as diagnostics for the physical origin of the luminous radio emission seen in the late phase of SNe.
For example, the existence of a massive CSM shell is supported in case the SN experiences the transformation of its morphology from a stripped-envelope SN to an interacting SN in the optical wavelength \citep[e.g.,][]{2017ApJ...835..140M}. In this case, X-ray emission is also expected to follow.
We suggest that multi-band observations during the radio re-brightening phase would be the unique diagnostics for the physical mechanism producing the double peak in the radio emission.

Furthermore, we expect that the multi-band radio observations to measure the spectral index can potentially distinguish the physical origin of the peaks. As mentioned in Section \ref{sec:demonstration}, all of the light curve peaks treated in this study appear in different ways with each other. The measured spectral index can diagnose the morphology of the synchrotron spectra (see Figure \ref{fig:concept}) so that we could examine the occurrence of the regime transition of the electron SED.

\subsection{Implications for shock acceleration physics}\label{subsec:microphysics}

We proposed that a double-peaked radio light curve emerges when $\epsilon_e$ is high and $f_e$ is low. This parameter set leads to the high minimum Lorentz factor of the injected electrons ($\gamma_m$). For example, $\gamma_m$ in the double-peaked light curve model in Section \ref{sec:demonstration} is as high as $\sim 200$ at $t=t_\ast$. The model fitted to SN 2007bg in Section \ref{subsec:observation} has even higher $\gamma_m \sim 1000$ at $t=t_\ast$. Such situations are interpreted as that all of the electrons are highly relativistic already at the moment of the injection to the diffusive shock acceleration process.
Then the SED of non-thermal electrons would not smoothly connect to the SED of thermal electrons.


{In the standard theory of diffusive shock acceleration (DSA), it is considered that electrons are injected at a low energy where the DSA is inefficient. Some of them eventually attain high Lorentz factor once the gyro radius exceeds the geometrical thickness of the shock \citep[e.g.,][]{1978MNRAS.182..147B, 1978MNRAS.182..443B, 1978ApJ...221L..29B, 1983RPPh...46..973D}. This picture indicates that only the electrons with the Lorentz factor of $\gamma_0 \sim (m_p/m_e)(V_{\rm sh}/c)\sim 200 (V_{\rm sh}/0.1c)$ would be involved in the DSA, and then most of the relativistic electron population reside in the Lorentz factor below $\gamma_0$ \citep[see also][]{2013ApJ...762L..24M}. Therefore, following this conventional picture, we would expect $\gamma_m < \gamma_0$, as is opposite to  the high values of $\gamma_m$ as the condition required for the double-peaked light curve. However, there is indeed no established consensus on the property of electron acceleration, and thus the above-mentioned picture may be over-simplified. As a future work, it is required to verify whether the high value of $\gamma_m$ can be realized from the physics of electron acceleration mechanism, and how this can be translated into the values of $\epsilon_e$ and $f_e$ in the phenomenological treatment. }

We also note that the properties of the low-energy side of the non-thermal electrons' SED in radio SNe have been studied in only a few previous studies. \cite{2015ApJ...805..164N} suggested the fitting formula taking into account the minimum energy of non-thermal electrons, indicating the overestimate of the explosion energy and the mass-loss rate of the SN progenitor. \cite{2013ApJ...762L..24M} argued that low-frequency radio emission from SNe can be used to study the so-called electron injection problem, by clarifying that the electrons responsible for the low-frequency radio emission are not energetic enough to be involved in the diffusive shock acceleration.
Our model has placed emphasis on the evaluation of $\gamma_m$, and thus may offer another clue to examine the properties of the electron SEDs at the low-energy side.

\subsection{Comparison with GRB afterglows}\label{subsec:afterglow}
GRB afterglow is synchrotron emission produced in the interaction region between the GRB jet and the interstellar medium, and we can observe it in the wide range from radio, optical, to X-ray \citep{1998ApJ...497L..17S, 2004RvMP...76.1143P, 2006RPPh...69.2259M, 2020ApJ...896..166R}.
The radiation mechanism is common to radio SNe except that the ejected material is relativistic in GRBs.
\cite{1998ApJ...497L..17S} constructed the standard model for GRB afterglows taking into account the regime transition of the electron SED.
However, the light curve suggested therein is characterized by a single peak.
This is attributed to the fact that the model for GRB afterglows does not incorporate synchrotron self-absorption.
The GRB system would be optically thin to synchrotron self-absorption because the external ambient medium is supposed to be a low-density interstellar medium. Even if we consider the stellar wind, the density can be low due to the large radius responsible for the radiation.
On the other hand, the CSM around the SN progenitor is characterized by the material released from the SN progenitor, which is denser by orders of magnitudes than the typical interstellar medium. Such a dense environment makes synchrotron self-absorption important, and thus, we need to take it into consideration in the modeling of radio SNe.

Another issue is related to the description of the radio luminosity.
The formulae used in GRB afterglows assume the whole shape of the synchrotron spectrum by the broken power-law distribution.
This cannot take into consideration the time variation of the integral interval in the synchrotron emissivity, but it has a critical role in reproducing the local minimum at the transition of the electron SED's regime (see Appendix \ref{app:derivation}).
We suggest that the modeling of synchrotron emission around the regime transition requires careful treatments of the time evolution of the characteristic Lorentz factors and the subsequent integrals in synchrotron quantities, even for GRB afterglow modelings.

In addition, there may be a cause in the hydrodynamics setup; in radio SNe, the self-similar solution in the non-relativistic regime is often assumed \citep{1982ApJ...258..790C}, while in GRB afterglows the solution for the relativistic blast wave is frequently adopted \citep{1976PhFl...19.1130B}. These two solutions have different time dependencies from each other and also produce the difference in the description of the resultant synchrotron quantities.

{
\subsection{Weakening the dip feature}\label{subsec:dip_treatment}
In this section, we mention two possible treatments that could weaken the dip feature in the double-peaked light curve. One is the assumption on the function of the electron SED. We have considered the broken power-law distribution {for simplicity}, where the bending points are neither smoothly connected nor cut off at the characteristic Lorentz factors (i.e., $\gamma_m$ and $\gamma_c$). This non-smooth profile may be important for characterizing the double-peaked feature in the radio light curve. However, if we rely on numerical simulations to construct the model for the electron SED \citep[e.g.,][]{2023MNRAS.524.6004W}, the smooth profile should appear in the electron SED, which make the dip feature in the radio light curve more inconspicuous.}

{Another effect is the difference in the arrival time of the radiation from the SN-CSM interaction system \citep[e.g.,][]{2019ApJ...870...38S}. Light from the foreground for the observer reaches us earlier than light from the background, and we can observe the light curve as a blend of the emission coming from each part of the SN-CSM interaction region. As we have assumed a spherically symmetric configuration for the hydrodynamical structure, we expect to observe a superposition of time-shifted light curves suggested in this paper. While this superposition can lead to the weakening of the dip feature in the radio light curve, its impact is expected to be relatively minor because the timescale of the light curve blending is determined by the ratio of $V_{\rm sh}/c$, smaller than unity. We also suppose that the effect of the arrival time difference is important even when we consider radio emission from SNe on the assumption of an inhomogeneous or an aspherical CSM structure.}

\section{Summary}\label{sec:summary}

In this study, we construct the model for calculating synchrotron emission induced by the interaction between SN ejecta and the CSM, {where an electron SED is given by} a broken power-law distribution. Therein two critical Lorentz factors become important; $\gamma_m$, the minimum Lorentz factor of the injected electrons, and $\gamma_c$, the Lorentz factor at which synchrotron (and inverse Compton) cooling rate balances with adiabatic cooling rate. Depending on the inequality between $\gamma_m$ and $\gamma_c$, the electron SED can take either the fast cooling or the slow cooling regime.
This method has been usually employed in the context of the modeling for GRB afterglows, and we apply the framework to radio SNe in a similar way.

This generalization of radio SN modeling leads to the finding that an SN can show not only a usual single-peaked but also a double-peaked light curve in the radio synchrotron emission even for single power-law CSM distribution, depending on the energy and number density of non-thermal electrons; this is a new regime that was not recognized previously. The first peak is characterized by $\tau_{\rm syn}=1$ during the fast cooling regime, while the second peak is formed at the moment of $\gamma_{\rm obs} = \gamma_m$ after the transition from the fast cooling to the slow cooling regime.
The condition for the emergence of the double-peaked radio light curve is that the system should become optically thin at the observational frequency during the fast cooling regime ($t_1 < t_\ast$); otherwise, only a single peak appears. Furthermore, to capture the double peak feature observationally, the transition timescale from fast cooling to slow cooling regimes ($t_\ast$) should be comparable to typical observational timescales.
The combination of high $\epsilon_e$ and low $f_e$ is likely to satisfy these conditions. This corresponds to the situation where the non-thermal electrons' SED is separated from thermal electrons' component. A broad-lined Type Ic SN 2007bg would be a possible double-peaked radio SN that can be fitted by our model. Our model can serve as potential diagnostics for plasma physics related to shock acceleration mechanism in SNe.

\section*{Acknowledgements}

The authors acknowledge Kohta Murase, Kaori Obayashi, Kenta Hotokezaka, Akihiro Suzuki, Toshikazu Shigeyama, Christopher Irwin, Yudai Suwa, {and the referee} for their fruitful comments.
This study is supported by the Japan Society for the Promotion of Science (JSPS) KAKENHI grant No. 20H01904, 21J12145 (TM), 22K14028, 23H04899 (SSK), 20H00174 (KM), 19H00694, 20H00158, 20H00179, 21H04997, 23H00127, 23H04894, 23H05432 (MT). {This research is supported by the National Science and Technology Council, Taiwan under grant No. MOST 110-2112-M-001-068-MY3 and the Academia Sinica, Taiwan under a career development award under grant No. AS-CDA-111-M04.} SSK acknowledges the support by the Tohoku Initiative for Fostering Global Researchers for Interdisciplinary Sciences (TI-FRIS) of MEXT's Strategic Professional Development Program for Young Researchers. MT acknowledges the support by the Japan Science and Technology Agency (JST) Fusion Oriented Research for Disruptive Science and Technology (FOREST) Program grant No. JPMJFR212Y. A part of the computations shown in this paper has been carried out in the computational cluster Yukawa-21 equipped in Yukawa Institute for Theoretical Physics.

\restartappendixnumbering
\appendix
\section{Proportional relationships}\label{app:proportional}
For completeness, we show the proportional dependencies of the quantities on SN age, CSM density, and shock acceleration efficiencies. They help us understand the behavior of the electron SED and the resultant radio light curve.
\begin{eqnarray}
R_{\rm sh} &\propto& \tilde{A}_\ast^{\frac{-1}{n-s}} t^m, \\
V_{\rm sh} &\propto& \tilde{A}_\ast^{\frac{-1}{n-s}} t^{m-1}, \\
\rho_{\rm sh} V_{\rm sh}^2 &\propto& \tilde{A}_\ast^{\frac{n-2}{n-s}} t^{m(2-s)-2}, \\
B &\propto& (\epsilon_B \rho_{\rm sh} V_{\rm sh}^2)^{1/2} \propto \epsilon_B^{\frac{1}{2}} \tilde{A}_\ast^{\frac{n-2}{2(n-s)}} t^{m(2-s)/2-1}, \\
\gamma_m &\propto& \epsilon_e f_e^{-1}V_{\rm sh}^2 \propto \epsilon_e f_e^{-1} \tilde{A}_\ast^{\frac{-2}{n-s}} t^{2(m-1)}, \\
\gamma_c &\propto& t^{-1} B^{-2} \propto \epsilon_B^{-1} \tilde{A}_\ast^{\frac{2-n}{n-s}} t^{-m(2-s)+1}, \\
\gamma_{\rm obs} &\propto& B^{-1/2} \propto \epsilon_B^{-1/4} \tilde{A}_\ast^{\frac{-n+2}{n-s}} t^{-m(2-s)/4+1/2}, \\
C &\propto& \epsilon_e \rho_{\rm sh} V_{\rm sh}^2 \gamma_m^{p-2} V_{\rm sh} \Delta R_{\rm sh}^{-1} \propto \epsilon_e^{p-1} f_e^{2-p} \tilde{A}_\ast^{\frac{n-2}{n-s} - \frac{2(p-2)}{n-s}} t^{m(2-s)-3+2(m-1)(p-2)}.
\end{eqnarray}

\section{Analytical descriptions on the radio luminosity}\label{app:derivation}
In this section, we derive the analytic formula {describing double-peaked radio light curves.}
The critical points are that (1) we should not adopt the $\delta$-function approximation for the synchrotron emission profile in Equation (\ref{eq:Psyn}) but should take the width of the profile into consideration, and (2) we {need to take care of the treatments on the upper and lower bounds of the integrals in Equation (\ref{eq:jsyn}).}

Let us assume the optically thin {situation at the observational frequency ($\nu$)} and start the integral of the emissivity (Equation (\ref{eq:jsyn})). To avoid the loss of the generality, we consider a single power-law distribution for non-thermal electrons given as $N(\gamma) = N_{\rm coeff} \gamma^{-k}$ with the domain of Lorentz factor defined as $\gamma_a < \gamma < \gamma_b$. If we assume a $\delta$-function profile for the synchrotron emission, $G(x) = g_\delta \delta(x-1)$, then the emissivity would become
\begin{eqnarray}
j_{\rm syn} = \frac{\sqrt{3} e^3 B}{8\pi m_e c^2} \left(\frac{\nu}{\nu_{\rm syn}(\gamma=1)}\right)^{(1-k)/2}\times \int_{x_b}^{x_a} N_{\rm coeff} g_\delta \delta(x-1) x^{(k-3)/2}dx,
\end{eqnarray}
where $x=\nu/\nu_{\rm syn}(\gamma)$ is defined, and $x_a$ and $x_b$ are the forms of $x$ in which $\gamma_a$ and $\gamma_b$ are substituted, respectively. $x_a$ and $x_b$ will be applied to one of $x_c = \nu/\nu_{\rm syn}(\gamma_c)$, $x_m = \nu/\nu_{\rm syn}(\gamma_m)$, {or $x_M = \nu/\nu_{\rm syn}(\gamma_M)$} depending on the {function of the electron SED}. The above integration cannot be appropriately evaluated unless $x_b < 1 < x_a$, equivalent to $\gamma_a < \gamma_{\rm obs} < \gamma_b$. This is not physically reasonable, because even in the case of $\gamma_{\rm obs} < \min(\gamma_m, \gamma_c)$, electrons with Lorentz factors lying around $\min(\gamma_m, \gamma_c)$ mainly covers the radio luminosity at the observed frequency with its spectrum proportional to $\nu^{1/3}$.

To derive the reasonable formula of the emissivity, we need to adopt the synchrotron profile with a finite width in Equation (\ref{eq:Psyn}). \cite{2010PhRvD..82d3002A} proposed the strict expression of $G(x)$ without using the integral but with the modified Bessel functions with orders of $1/3$ and $2/3$ as follows:
\begin{eqnarray}
G(x) = \frac{x}{20}\left[(8+3x^2)(\kappa_{1/3})^2 + x\kappa_{2/3}(2\kappa_{1/3}-3x\kappa_{2/3})\right],
\end{eqnarray}
where $\kappa_d = K_d(x/2)$ is defined. To obtain analytic expressions we consider the approximated formulae for the modified Bessel functions in the limit of $x\rightarrow0$ and $x\rightarrow\infty$, which can be described as follows:
\begin{eqnarray}
\kappa_{d} = 
\left\{
\begin{array}{cc}
  {\displaystyle \frac{4^d \pi}{\sqrt{3}\Gamma(1-d)}x^{-d}}   & (x\rightarrow 0)  \\
  {\displaystyle \sqrt{\frac{\pi}{x}}\left(1+\frac{1}{x}\frac{\Gamma(d+3/2)}{\Gamma(d-1/2)}+\frac{1}{x^2}\frac{\Gamma(d+5/2)}{\Gamma(d-3/2)}\right)\exp(-x/2)} &   (x\rightarrow \infty),
\end{array}
\right.
\end{eqnarray}
where $\Gamma(d)$ is the Gamma function \citep{1965hmfw.book.....A}. As for the limiting case of $x\rightarrow \infty$, we have expanded the modified Bessel functions up to the second order to obtain the approximated expression with the minimum order of $G(x)$. We obtain $G(x) \simeq 1.808 x^{1/3}$ in the limit of $x\rightarrow0$, while we can deduce $G(x) \simeq (179\pi/288)e^{-x} \simeq 1.952e^{-x}$ in the limit of $x\rightarrow\infty$. Incorporating these two limiting cases, we assume 
\begin{eqnarray}
G(x) = gx^{1/3} {\exp(-x)},\label{eq:Gapprox}
\end{eqnarray}
where $g$ is a coefficient factor of $1.5\sim2$.
The choice of $g$ would affect the normalization of the radio luminosity but not alter the qualitative behaviors of the light curve.
Equation (\ref{eq:Gapprox}) is compatible with the formula written in \cite{2008FrPhC...3..306F}.

So far we have considered the non-thermal electrons' SED given by a single power-law function. However, what we have discussed in this study is the broken power-law distribution realized in the fast cooling or slow cooling regime. Therefore, we need to take a summation of the two power-law distributions in the electron spectrum.
When we consider the fast cooling case, the emissivity can be written as
\begin{eqnarray}\label{eq:jsyn_integral_fast}
j_{\rm syn} \simeq \frac{\sqrt{3}e^3 Bg}{8\pi m_e c^2}
\left\{\left(\frac{\nu}{\nu_{\rm syn}(\gamma=1)}\right)^{-1/2}\int_{x_m}^{x_c}\frac{Ct_{\rm syn}(\gamma=1)\gamma_m^{1-p}}{p-1} x^{-1/6}\exp(-x)dx\right. \nonumber \\
\left. + \left(\frac{\nu}{\nu_{\rm syn}(\gamma=1)}\right)^{-p/2} \int_{x_M}^{x_m} \frac{Ct_{\rm syn}(\gamma=1)}{p-1} x^{(3p-4)/6} \exp(-x) dx \right\}.
\end{eqnarray}
{As discussed in Section \ref{subsubsec:gamma_SED}, $\gamma_M$ is orders of magnitude greater than $\gamma_m$ so that we can approximate $x_M\simeq 0$.}
On the other hand, we are considering time-variable $\gamma_m$ and $\gamma_c$ changing their size relationship with each other.
Then, we need to leave the upper and lower bound of the above integral finite, not to approximate $x_m$ and $x_c$ as zero.
Instead, it is permitted to approximate the exponential factor in the integral as unity, $\exp(-x) \sim1$, because $x_m$ and $x_c$ are now {sufficiently} smaller than 1. Given the above conditions, some algebra allows us to derive the following expression,
\begin{eqnarray}\label{eq:emissivity}
j_{\rm syn} \simeq \frac{3\sqrt{3}e^3BCt_{\rm syn}(\gamma=1)\gamma_m^{1-p}g}{20\pi m_e c^2(p-1)}\left(\frac{\nu}{\nu_{\rm syn}(\gamma=1)}\right)^{1/3} \left(\gamma_c^{-5/3}-\frac{3p-3}{3p+2}\gamma_m^{-5/3}\right).
\end{eqnarray}
We can see that the time evolution of the emissivity consists of two components; one is the term dependent on $\gamma_c^{-5/3}$, accounting for the effect that the radio luminosity is covered by electrons with Lorentz factors $\gamma\sim\gamma_c$. The other is the term proportional to $\gamma_m^{-5/3}$, meaning that the decrease in $\gamma_m$ leads to the reduction of the population of emitter electrons.

From Equation (\ref{eq:emissivity}), we can infer the origin of the local minimum of the double-peaked radio light curve. Multiplying $4\pi^2 R_{\rm sh}^2 \Delta R_{\rm sh}$ leads to the formula of the radio luminosity, which can be described as follows:
\begin{eqnarray}
L_{\nu} &\simeq& \frac{3\sqrt{3}\pi g e^3 R_{\rm sh}^2 \Delta R_{\rm sh} BC t_{\rm syn}(\gamma=1)\gamma_m^{1-p}}{5m_e c^2(p-1)}\left(\frac{\nu}{\nu_{\rm syn}(\gamma=1)}\right)^{\frac{1}{3}} \left(\gamma_c^{-\frac{5}{3}} - \frac{3p-3}{3p+2} \gamma_m^{-\frac{5}{3}}\right) \\
& \equiv& {\displaystyle \nu^{\frac{1}{3}}\left(L_{1,c} t^{\frac{15m-6ms-4}{3}} - L_{1,m} t^{\frac{-5m-ms+11}{3}}\right)},\label{eq:Lnu_fast}
\end{eqnarray}
where $L_{1,c}$ and $L_{1,m}$ are the coefficients independent of $t$ and $\nu$, originating from $\gamma_c^{-5/3}$ and $\gamma_m^{-5/3}$, respectively.
As for the parameter set in Section \ref{sec:demonstration}, $s=2$ and $m=0.8$, and thus, the index of the first term is ${-8/15} \simeq {-0.53}$. Because $\gamma_c$ increases with time, the optically thin emission would decline with time intrinsically as long as $\gamma_{\rm obs} < \gamma_c$.
Furthermore, the second term depends on $t^{1.8}$, indicating that the decrease of $\gamma_m$ brings about the suppression of the radio luminosity. In the end, the system reaches the situation in which $\gamma_c$ and $\gamma_m$ coincide with each other.
This moment can be observed as the local minimum of the radio luminosity.

{After the transition from fast cooling to slow cooling regime, $\gamma_m < \gamma_c$ is realized, as well as the shape of the electron SED changes. As for the epoch before reaching the second peak, we can write down the emissivity as follows:
\begin{eqnarray}
j_{\rm syn} &\simeq& \frac{\sqrt{3}e^3 Bg}{8\pi m_e c^2}
\left\{\left(\frac{\nu}{\nu_{\rm syn}(\gamma=1)}\right)^{(1-p)/2}\int_{x_c}^{x_m}\frac{Ct_{\rm ad}}{p-1} x^{(3p-7)/6}\exp(-x)dx\right. \nonumber\label{eq:j_syn_slow1} \\
&& \qquad \left. + \left(\frac{\nu}{\nu_{\rm syn}(\gamma=1)}\right)^{-p/2} \int_{x_M}^{x_c} \frac{Ct_{\rm syn}(\gamma=1)}{p-1} x^{(3p-4)/6} \exp(-x) dx \right\} \\
&\simeq&\frac{3\sqrt{3}e^3BCt_{\rm ad}g}{4(3p-1)\pi m_e c^2(p-1)}\left(\frac{\nu}{\nu_{\rm syn}(\gamma=1)}\right)^{1/3}\left(\gamma_m^{-p+1/3} -\frac{3}{3p+2}\gamma_c^{-p+1/3}\right),\label{eq:b10}
\end{eqnarray}
where $\exp(-x)\sim 1$ is taken into consideration because $\gamma_{\rm obs}>\gamma_m$ is still satisfied immediately after the regime transition. Then the radio luminosity has dependencies on time and observational frequency as follows:
\begin{eqnarray}
L_\nu \equiv \nu^{\frac{1}{3}}(L_{2,m}t^{\frac{7m-4ms+2}{3}} - L_{2,c}t^{m-ms+\frac{5}{3}+p(4m-ms-3)}),\label{eq:Lnu_slow1}
\end{eqnarray}
where $L_{2,m}$ and $L_{2,c}$ are the coefficients independent of $t$ and $\nu$. The first and second terms originate from $\gamma_m^{-p+1/3}$ and $\gamma_c^{-p+1/3}$ in Equation (\ref{eq:b10}), and are proportional to $t^{0.4}$ and $t^{-3.3}$ for the parameter set in Section \ref{sec:demonstration}, respectively. This indicates that the second brightening in the double-peaked radio light curve is attributed to the time evolution of $\gamma_m$ approaching $\gamma_{\rm obs}$, while the contribution from electrons with $\gamma\sim\gamma_c$ can be neglected.
}

{Finally we describe the behavior of the radio luminosity after the second peak. During this phase, $\gamma_m$ decreases below $\gamma_{\rm obs}$ so that the approximation of $\exp(-x)\sim 1$ becomes invalid. Instead, we could approximate $x_m\sim\infty$ and $x_c\sim0$ in the limit of $t\rightarrow \infty$ as an asymptotic form in Equation (\ref{eq:j_syn_slow1}). This operation neglects the contribution from the SED component in $\gamma > \gamma_c$. Then
\begin{eqnarray}
j_{\rm syn} &\simeq& \frac{\sqrt{3}e^3 Bg}{8\pi m_e c^2}
\left(\frac{\nu}{\nu_{\rm syn}(\gamma=1)}\right)^{(1-p)/2}\int_{0}^{\infty}\frac{Ct_{\rm ad}}{p-1} x^{(3p-7)/6}\exp(-x)dx \\
&=& \frac{\sqrt{3}e^3BCt_{\rm ad}g}{8\pi m_e c^2(p-1)} \Gamma\left(\frac{3p-1}{6}\right)\left(\frac{\nu}{\nu_{\rm syn}(\gamma=1)}\right)^{(1-p)/2},
\end{eqnarray}
can be deduced. The radio luminosity has dependencies on time and observational frequency as follows:
\begin{eqnarray}
L_\nu \propto \nu^{\frac{(1-p)}{2}}t^{-m-\frac{5}{4}s+4+p(2m-2-\frac{s}{4})},\label{eq:Lnu_slow2}
\end{eqnarray}
which is proportional to $t^{-2}$ for the parameter set in Section \ref{sec:demonstration}. In summary, we can find out from Equations (\ref{eq:Lnu_fast}), (\ref{eq:Lnu_slow1}), and (\ref{eq:Lnu_slow2}) the time dependence of the radio luminosity decaying from the first peak, re-brightening in the second peak, and declining from the second peak, respectively. These behaviors, in addition to the effect of synchrotron self-absorption in the early phase, would shape the double-peaked radio light curve.}

\bibliographystyle{aasjournal}
\bibliography{manuscript}{}

\end{document}